\documentclass[ALICE,manyauthors]{cernphprep}

\usepackage[comma,square,numbers,sort&compress]{natbib}
\usepackage{hyperref}
\usepackage{lineno}

\newcommand{\GeVc}{\ensuremath{\mathrm{GeV}\kern-0.05em/\kern-0.02em c}}
\newcommand{\GeVcSq}{\ensuremath{\mathrm{GeV}\kern-0.05em/\kern-0.02em c^2}}

\begin{document}%

\begin{titlepage}
\PHyear{2018}
\PHnumber{098}      
\PHdate{8 May}  
%

\title{Measurement of the inclusive J/$\psi$ polarization at forward rapidity \\ in pp collisions at $\mathbf{\sqrt{s} = 8}$~TeV}
\ShortTitle{Forward J/$\psi$ polarization in pp at $\sqrt{s} = 8$~TeV}   

\Collaboration{ALICE Collaboration\thanks{See Appendix~\ref{app:collab} for the list of collaboration members}}
\ShortAuthor{ALICE Collaboration} 

\begin{abstract}
We report on the measurement of the inclusive J/$\psi$ polarization parameters in pp collisions at a center of mass energy $\sqrt{s} = 8$~TeV with the ALICE  detector at the LHC. 
The analysis is based on a data sample corresponding to an integrated luminosity of 1.23~pb$^{-1}$. 
J/$\psi$ resonances are reconstructed in their di-muon decay channel in the rapidity interval $2.5 < y < 4.0$ and over the transverse-momentum interval $2 < p_{\rm T} < 15$~\GeVc.
The three polarization parameters ($\lambda_\theta$, $\lambda_\varphi$, $\lambda_{\theta\varphi}$) are measured as a function of $p_{\rm T}$ both in the helicity and Collins-Soper reference frames.
The measured J/$\psi$ polarization parameters are found to be compatible with zero within uncertainties, contrary to expectations from all available predictions.
The results are compared with the measurement in pp collisions at $\sqrt{s} = 7$~TeV.
\end{abstract}
\end{titlepage}
\setcounter{page}{2}

\section{Introduction}
\label{sec:intro}

More than 40 years after the J/$\psi$ discovery, its production mechanism in hadronic collisions remains an open issue~\cite{brambilla:2010cs}. 
Quarkonia states constitute an important test bench for the study of Quantum ChromoDynamics (QCD) both in the vacuum and in high-energy density environments, as those produced in heavy-ion collisions, where the creation of the Quark--Gluon Plasma (QGP) is observed~\cite{Andronic:2015wma}.
Consequently, the understanding of the J/$\psi$ production mechanism is an important scientific question in the sense that it addresses basic concepts of QCD, the theory of the strong interaction, and its application to heavy-ion collisions allows the characterisation of the QGP properties created in the laboratory.

Different theoretical models have been developed in an attempt to describe the whole production mechanism from partonic interaction to heavy-quark pair (Q$\overline{\rm Q}$) hadronisation in quarkonia. 
All approaches are based on the factorisation hypothesis between hard and soft scales.
First phenomenological attempts (e.g.~the Color Evaporation Model~\cite{fritzsch:1977ay}) have been replaced by a rigorous effective field theory, the Non-Relativistic QCD (NRQCD)~\cite{Bodwin:1994jh}. 
In this framework, two models can be derived according to the sub-processes taken into account: the Color-Singlet Model (CSM) ~\cite{Einhorn:1975ua, Lansberg:2008gk} and the Color-Octet Mechanism (COM)~\cite{Bodwin:1994jh}.
The CSM assumes no evolution of the quantum color-singlet state between the Q$\overline{\rm Q}$ production and the quarkonium formation, with a wave function computed at zero Q$\overline{\rm Q}$ separation, i.e.\ without any free parameter.
The COM introduces Long-Distance Matrix Elements (LDMEs) for the hadronisation probability in a quarkonium state. 
The LDMEs are free parameters of the theory which must be fixed from experimental data.

Recent measurements at the LHC confirm that color-octet terms are crucial for a good description of the J/$\psi$ and $\psi$(2S) differential production cross sections~\cite{Acharya:2017hjh}. 
However, the failure in predicting the $\eta_{\rm c}$ production cross section~\cite{Aaij:2014bga,Butenschoen:2014dra} poses serious challenges to the NRQCD approach.

In this context, alternative measurements at different energies and in different rapidity regions can help to disentangle tensions between quarkonium measurements and the theoretical predictions.
One of the most relevant observables apart from the production cross section is the polarization of quarkonia.
The polarization of $\rm J^{\rm PC} = 1^{--}$ states like the J/$\psi$ is specified by three polarization parameters ($\lambda_\theta$, $\lambda_\varphi$, $\lambda_{\theta\varphi}$), which are a function of the three decay amplitudes with respect to the three angular momentum states.
The two cases ($\lambda_\theta = 1$, $\lambda_\varphi = 0$, $\lambda_{\theta\varphi} = 0$) and ($\lambda_\theta = -1$, $\lambda_\varphi = 0$, $\lambda_{\theta\varphi} = 0$) correspond to the so-called transverse and longitudinal polarizations, respectively.
Theoretical models at Next-to-Leading Order (NLO) predict strongly transverse-momentum dependent polarization states with a partial longitudinal polarization in the CSM and a partial transverse polarization when color-octet contributions are included in the NRQCD calculation~\cite{butenschoen:2012px}.

Experimentally, the polarization parameters can be determined in the quarkonium dilepton decay channel by studying the angular distribution ($W$) of the leptons in the quarkonium rest-frame~\cite{Faccioli:2010kd}: 
\begin{eqnarray}
\label{equ:W}
W(\cos\theta,\varphi) & \propto & \frac{1}{3+\lambda_\theta} \left[ 1 + \lambda_\theta \cos^2\theta + \lambda_\varphi \sin^2\theta \cos(2\varphi) + \lambda_{\theta\varphi} \sin(2\theta) \cos\varphi \right] \label{equ:0}
\end{eqnarray}
where $\theta$ and $\varphi$ are the polar and the azimuthal angles, respectively, defining the orientation of one lepton (for instance the negative one) in the quarkonium rest-frame with respect to a reference axis.
In the analysis presented here, the selected reference axes are: (i) the helicity axis corresponding to the quarkonium flight direction in the center-of-mass of the colliding beams, and (ii) the Collins-Soper axis defined by the direction of the relative velocity of the colliding beams in the quarkonium rest-frame.
In the following, the J/$\psi$ rest-frame associated to the helicity axis will be referred to as helicity (HX) frame and the one defined from the Collins-Soper axis will be called Collins-Soper (CS) frame.

Since the beginning of the LHC operations, the study of the J/$\psi$ polarization in pp collisions has been carried out at $\sqrt{s} = 7$~TeV both at midrapidity by the CMS~\cite{Chatrchyan:2013cla} experiment, and at forward rapidity by the ALICE~\cite{Abelev:2011md} and LHCb~\cite{Aaij:2013nlm} experiments. 
The midrapidity and forward rapidity results are complementary in terms of the explored transverse-momentum ($p_{\rm T}$) interval, which is $14 < p_{\rm T} < 70$~\GeVc\ for CMS, $2 < p_{\rm T} < 15$~\GeVc\ for LHCb and $2 < p_{\rm T} < 8$~\GeVc\ for ALICE.

In this paper we present the polarization measurement of inclusively-produced J/$\psi$ mesons in pp collisions at $\sqrt{s} = 8$~TeV in the transverse-momentum interval $2 < p_{\rm T} < 15$~\GeVc.
This is the first measurement of the J/$\psi$ polarization at this energy, and extends the $p_{\rm T}$ reach of the previous ALICE measurement at $\sqrt{s} = 7$~TeV~\cite{Abelev:2011md}.
The paper starts with a brief description of the experimental apparatus and the used data sample in Section~\ref{sec:apparatus}, followed by a description of the analysis in Section~\ref{sec:analysis}, including a discussion of the systematic uncertainties.
The results are presented in Section~\ref{sec:results} and compared with those obtained from $\sqrt{s} = 7$~TeV and with model calculations. 
Conclusions are finally drawn in Section~\ref{sec:conclusion}.

\section{Experimental apparatus and data sample}
\label{sec:apparatus}

The ALICE apparatus and its performance are described in detail in~\cite{Aamodt:2008zz} and ~\cite{Abelev:2014ffa}, respectively.
In this paper we focus on the two sub-detectors relevant for the analysis: the forward muon spectrometer~\cite{Aamodt:2011gj} and the first two layers of the Inner Tracking System (ITS)~\cite{Aamodt:2010aa}.

The muon spectrometer detects muons in the pseudorapidity range\footnote{Although the muon spectrometer covers negative pseudorapidities ($\eta$) in the ALICE reference frame, we use positive rapidity values when referring to the rapidity ($y$) of quarkonia states reconstructed via their di-muon decay channel.} $-4.0 < \eta < -2.5$. 
It consists of five tracking stations with two detection planes of multi-wire proportional chambers with cathode pad readout and two trigger stations, each comprising of two detection planes of resistive plate chambers. 
A set of absorbers completes the system, to decrease the hadronic background: the front-absorber (before the first tracking station) reduces the contamination of light hadron decays, a shield surrounding the beam pipe decreases the background from particles produced in the interaction at large pseudorapidity, and an iron wall shields the trigger stations from residual punch through.
The momenta of charged tracks are measured with the help of a 3~T$\cdot$m dipole magnet surrounding the third tracking station.

The ITS consists of six layers of silicon detectors with cylindrical geometry surrounding the beam pipe, with radii ranging from 3.9 to 43~cm from the beam axis. 
This analysis makes use of the two innermost layers that are equipped with Silicon Pixel Detectors (SPD) and cover the pseudorapidity ranges $|\eta| < 2$ and $|\eta| < 1.4$ for the first and the second layer, respectively. 
The SPD is used to reconstruct the position of the primary vertex of the collision. 

The data used for this analysis were collected in 2012.
The online event selection is based on the opposite-sign di-muon trigger, with a $p_{\rm T}$ threshold of about 1~\GeVc\ applied on each muon candidate.
This di-muon trigger runs in coincidence with the crossing of two beam bunches at the interaction point.
The data sample recorded with this trigger configuration is the same as in~\cite{Adam:2015rta} and corresponds to an integrated luminosity of about 1.23~pb$^{-1}$.

\section{Analysis}
\label{sec:analysis}
 
\paragraph{Track selection.}
The opposite-sign di-muon pair candidates are reconstructed with the following track selection criteria (see~\cite{Adam:2015rta} for details):
\begin{itemize}
\item
the track pseudorapidity must be in the range corresponding to the muon spectrometer acceptance $-4 < \eta < -2.5$,
\item 
the polar angle $\theta_{\rm abs}$ measured at the rear-end plane of the front absorber must be in the interval \\ $170 < \theta_{\rm abs} < 178$ degrees,
\item
the maximum allowed value for the $p$DCA variable, defined as the product of the total momentum $p$ of the track and its distance of closest approach DCA to the primary vertex in the transverse plane, must be less than $6 \times \sigma_{p{\rm DCA}}$, where the resolution $\sigma_{p{\rm DCA}}$ is 54~cm$\cdot$\GeVc\ for $170 < \theta_{\rm abs} < 177$ degrees and 80~cm$\cdot$\GeVc\ for $177 \le \theta_{\rm abs} < 178$ degrees, 
\item
each track reconstructed in the muon tracking system must match a track in the trigger system and in addition must pass the low-$p_{\rm T}$ trigger threshold of $\sim 1$~\GeVc .
\end{itemize}
Finally, each unlike-sign di-muon pair is required to be in the rapidity interval $2.5 < y < 4.0$.

\paragraph{J/$\psi$ polarization formalism.}
A polarization analysis performed by fitting for each $p_{\rm T}$ interval the two-dimensional angular distribution of Eq.~(\ref{equ:W}) requires a large reconstructed J/$\psi$ sample.
In the present analysis, given the limited statistics, the two-dimensional angular distribution is integrated over one angle at a time, to obtain the three following normalised one-dimensional distributions:
\begin{equation}
\label{equ:1}
W_1(\cos\theta) = \frac{3N}{2(3+\lambda_\theta)} \left[ 1 + \lambda_\theta \cos^2\theta \right] 
\end{equation}
\begin{equation}
\label{equ:2}
W_2(\varphi) = \frac{N}{2\pi} \left[ 1 + \frac{2\lambda_\varphi}{3+\lambda_\theta} \cos(2\varphi) \right] 
\end{equation}
\begin{equation}
\label{equ:3}
W_3(\widetilde{\varphi}) = \frac{N}{2\pi} \left[ 1 + \frac{\sqrt{2}\lambda_{\theta\varphi}}{3+\lambda_\theta} \cos\widetilde{\varphi} \right]
\end{equation}
with $\widetilde{\varphi} = \varphi - \frac{3}{4}\pi$ for $\cos\theta < 0$ and $\widetilde{\varphi} = \varphi - \frac{1}{4}\pi$ for $\cos\theta > 0$,
while $N$ corresponds to the normalisation factor common to the three distributions.

\paragraph{Analysis strategy.}
In order to extract the polarization parameters as a function of $p_{\rm T}$, the three angular distributions $W_1(\cos\theta)$, $W_2(\varphi)$ and $W_3(\widetilde{\varphi})$ are built by classifying the di-muon candidates in $\cos\theta$, $\varphi$ and $\widetilde{\varphi}$ intervals, respectively, for each $p_{\rm T}$ interval.
The raw number of J/$\psi$ mesons is extracted in each interval of $p_{\rm T}$ and angle via a fit of the corresponding invariant mass distribution. 
The fit is performed in the invariant mass range $2 < M_{\mu^+\mu^-} < 5$~\GeVcSq\ using a variable-width Gaussian function to describe the background shape and two extended Crystal Ball functions~\cite{ALICE-PUBLIC-2015-006} to describe the J/$\psi$ and $\psi$(2S) resonances.
The total number of J/$\psi$ in the analyzed data sample is about $50\,000$ in the transverse momentum range $2 < p_{\rm T} < 15$~\GeVc. 
The extracted raw yields are then corrected for the acceptance and efficiency of the detector ($A$$\times$$\epsilon$).

\begin{figure}[h]
\centering
\includegraphics[width=0.9\textwidth]{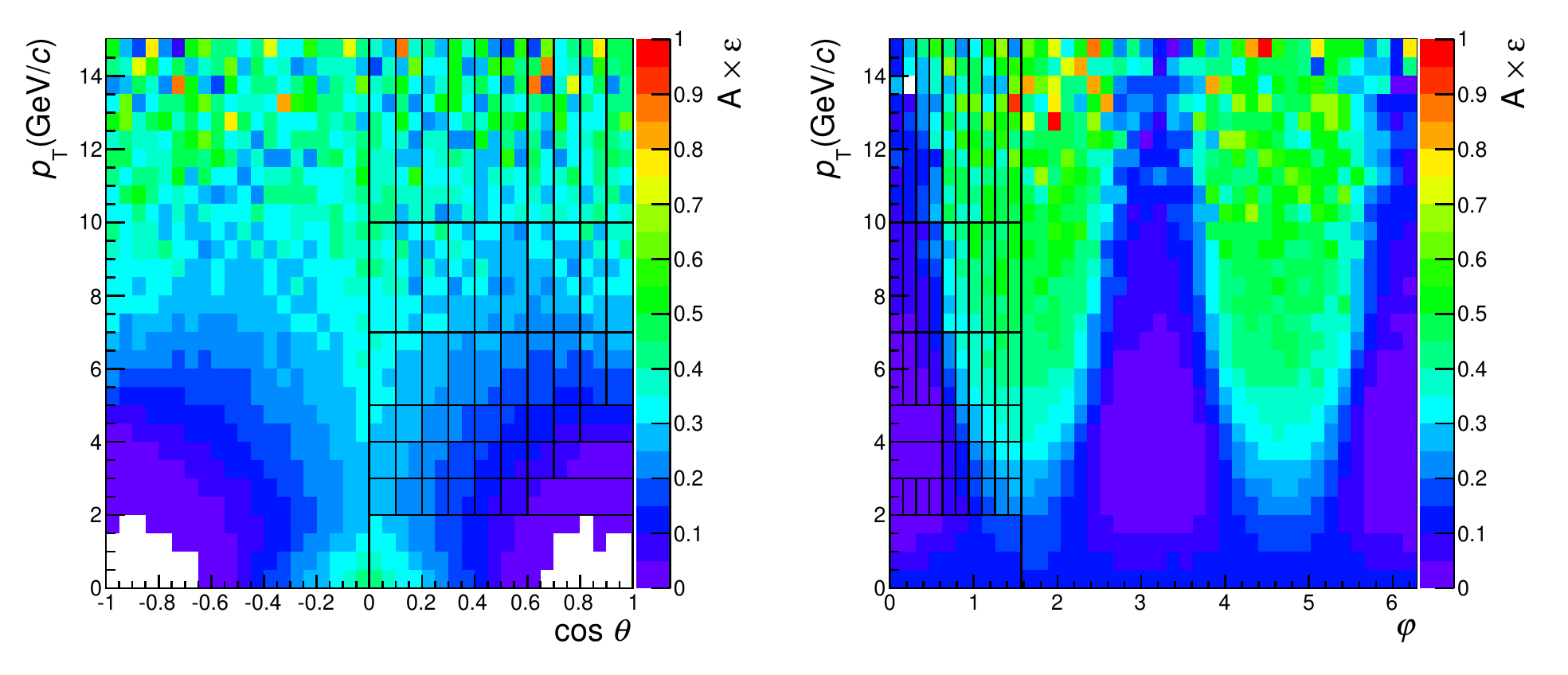}
\caption{($A$$\times$$\epsilon$) 2-D maps in the planes ($\cos\theta$, $p_{\rm T}$) (left) and ($\varphi$, $p_{\rm T}$) (right) in the Collins-Soper frame. The plots illustrate the symmetry with respect to $\cos\theta = 0$ (left) and with respect to $\varphi = \pi$ and $\pi/2$ (right), while the grid shows the binning used to build the $W_1(\cos\theta)$ and $W_2(\varphi)$ distributions in each $p_{\rm T}$ range.}
\label{fig:acc-eff}   
\end{figure}

\paragraph{Acceptance and efficiency evaluation.}
This is estimated with Monte Carlo (MC) simulations of unpolarized J/$\psi$ mesons with $p_{\rm T}$ and rapidity input distributions parameterized from the measured ones at the same energy~\cite{Adam:2015rta}.
Next, the J/$\psi$ mesons are forced to decay into $\mu^+ \mu^-$ pairs~\cite{Lange:2001uf}, including a fraction (5.4\%) of radiative decays $\mu^+ \mu^- \gamma$~\cite{Barberio:1990ms} in agreement with the prediction from~\cite{Spiridonov:2004mp}.
In the simulation, the particles are propagated through the ALICE apparatus using GEANT 3.21~\cite{Brun:1994aa} with a realistic description of the detector response.
The ($A$$\times$$\epsilon$) factor is calculated in each interval of $p_{\rm T}$ and angle as the ratio of reconstructed J/$\psi$ satisfying the selection criteria to the number of generated J/$\psi$ in the rapidity range $2.5 < y < 4.0$. 
As an example, Fig.~\ref{fig:acc-eff} (left) shows the ($A$$\times$$\epsilon$) map in the plane ($\cos\theta$, $p_{\rm T}$) for the CS frame.
A similar map is obtained in the HX frame, but with a vanishing ($A$$\times$$\epsilon$) in the interval $0.9 < |\cos\theta| < 1$ for $2 < p_{\rm T} < 15$~\GeVc.
The maps as a function of $\varphi$ and $\tilde{\varphi}$ in both frames do not exhibit any hole in the ($A$$\times$$\epsilon$), as illustrated in Fig.~\ref{fig:acc-eff} (right) in the plane ($\varphi$, $p_{\rm T}$) for the CS frame.
Due to the natural symmetry of the angular distributions the analysis is performed in the intervals $0 \le \cos\theta \le 1$, $0 \le \varphi \le \frac{\pi}{2}$ and $0 \le \widetilde{\varphi} \le \pi$.
The $p_{\rm T}$ interval explored in this analysis is constrained by a vanishing ($A$$\times$$\epsilon$) at low $p_{\rm T}$ and high $|\cos\theta|$, and by the limited statistics at high $p_{\rm T}$.
The angular distribution intervals for the analysis are defined in order to have a significance\footnote{The significance is defined as ${\cal S} = S/\sqrt{S+B}$ with $S$ the number of signal events and $B$ the number of background events in the mass range of $\pm 3 \sigma$ around the J/$\psi$ mass peak, $\sigma$ being the J/$\psi$ mass resolution.} larger than five.
The grid in Fig.~\ref{fig:acc-eff} shows the defined $p_{\rm T}$ ranges as well as the $\cos\theta$ (left plot) and $\varphi$ (right plot) intervals in the CS frame.

\begin{figure}[h]
\centering
\includegraphics[width=1.0\textwidth]{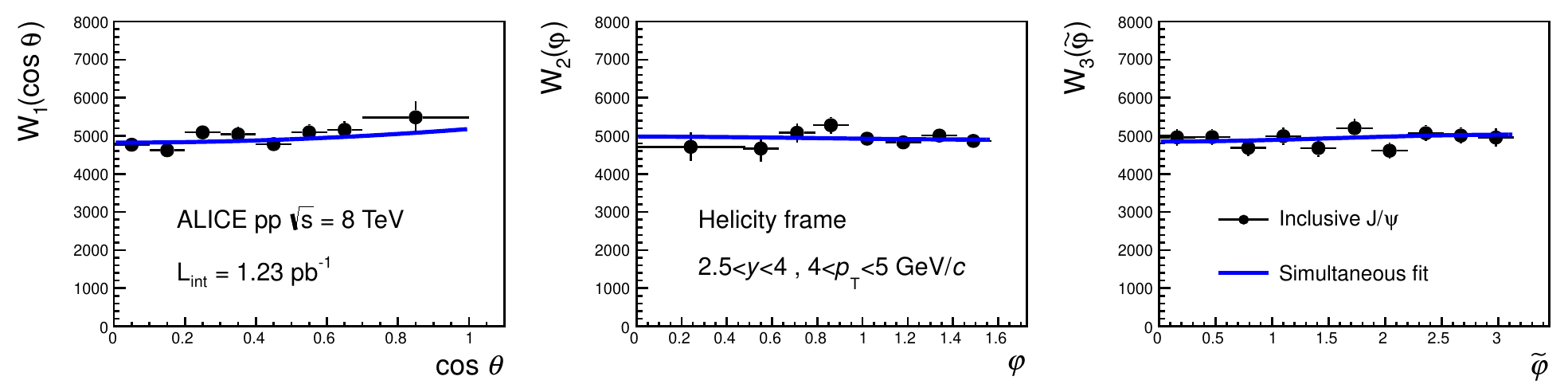}
\caption{Acceptance corrected angular distributions of J/$\psi$ reconstructed in the di-muon decay channel $W_1(\cos\theta)$, $W_2(\varphi)$ and $W_3(\widetilde{\varphi})$ in the helicity frame for the transverse momentum interval $4 < p_{\rm T} < 5$~\GeVc, together with  the results of the simultaneous fit (see text for details). Vertical bars correspond to statistical uncertainties.}
\label{fig:lambda_fit}   
\end{figure}

\paragraph{Extraction of the polarization parameters.}
After acceptance and efficiency correction of the number of reconstructed J/$\psi$ candidates, a simultaneous fit of the three angular distributions is performed by minimizing the following $\chi^2$-function for each $p_{\rm T}$ interval
\begin{eqnarray}
\chi^2 & = & \sum_{i=1}^{n_{\cos\theta}} \left(\frac{N_i^{\rm J/\psi} - W_1(\cos\theta\,;\,N,\lambda_\theta)}{\sigma_i}\right)^2 \nonumber\\ 
& & + \sum_{j=1}^{n_{\varphi}} \left(\frac{N_j^{\rm J/\psi} - W_2(\varphi\,;\,N,\lambda_\theta,\lambda_\varphi)}{\sigma_j}\right)^2 \nonumber\\ 
& & + \sum_{k=1}^{n_{\widetilde{\varphi}}} \left(\frac{N_k^{\rm J/\psi} - W_3(\widetilde{\varphi}\,;\,N,\lambda_\theta,\lambda_{\theta\varphi})}{\sigma_k}\right)^2 \label{equ:chi2}
\end{eqnarray}
with four free parameters: the normalization factor $N$ common to the three distributions and the three polarization parameters ($\lambda_\theta$, $\lambda_\varphi$, $\lambda_{\theta\varphi}$).
In this expression, $N_{i,j,k}^{\rm J/\psi}$ and $\sigma_{i,j,k}$ are the corrected numbers of J/$\psi$ and their associated statistical uncertainties in the $i^{\rm th}$, $j^{\rm th}$ and $k^{\rm th}$ bins of the angular distributions $W_1(\cos\theta)$, $W_2(\varphi)$ and $W_3(\widetilde{\varphi})$, with a total number of bins $n_{\cos\theta}$, $n_{\varphi}$ and $n_{\widetilde{\varphi}}$, respectively.
Figure~\ref{fig:lambda_fit} illustrates the fit results of the angular distributions in the HX frame for the transverse-momentum range $4 < p_{\rm T} < 5$~\GeVc\ (similar fits are obtained in all $p_{\rm T}$ intervals and in both frames).

\paragraph{Systematic uncertainty evaluation.}
The J/$\psi$ signal is extracted using five different fitting approaches. 
The initial approach of the invariant mass fit presented above is varied in the following way.
The range of the fit is increased to $1.5 < M_{\mu^+\mu^-} < 6$~\GeVcSq\ or decreased to $2.2 < M_{\mu^+\mu^-} < 4.5$~\GeVcSq. 
The product of a Gaussian and an exponential is used as an alternative background shape, and finally the two Crystal Ball functions are replaced by the function used by the NA60 Collaboration~\cite{ALICE-PUBLIC-2015-006}.
For each approach the analysis is repeated and the polarization parameters are determined. 
The final values of the polarization parameters $\lambda_\alpha$ (with $\alpha = \theta$ or $\varphi$ or $\theta\varphi$) correspond to the mean values $\overline{\lambda}_\alpha$ of the five sets of results, and the associated statistical uncertainties are the mean values of the statistical uncertainties returned by each fit.
The systematic uncertainties on the $\lambda_\alpha$ parameters due to the signal extraction are the sum of the quadratic difference of each configuration result with respect to the mean values.
The uncertainties range from 0.012 to 0.108 (see Table~\ref{tab:errors}) with the biggest effect observed on $\lambda_\varphi$ in the HX frame.
 
An exhaustive investigation of potential biases in the ($A$$\times$$\epsilon$) map is carried out.
Firstly, the input distributions of the J/$\psi$ in the MC simulation are modified by: (i) varying the $p_{\rm T}$ and (ii) the $y$ shapes of the J/$\psi$ parameterization within the uncertainties of the measured cross sections~\cite{Adam:2015rta}, (iii) removing the radiative decay part, (iv) varying the $\lambda_\theta$ parameter in the range $-0.2 < \lambda_\theta < 0.2$, corresponding to 1-sigma deviation of the measured value of $\lambda_\theta$ in the HX frame on average over the whole $p_{\rm T}$ interval.
The four corresponding uncertainties on the $\lambda_\alpha$ are summed quadratically to get the total simulation input uncertainties ranging from 0.004 to 0.175.
This is the main systematic uncertainty for the $\lambda_\theta$ parameter, dominated by the variation of its input value in simulation, especially at low $p_{\rm T}$.
Secondly, any uncertainty in the simulation of the trigger threshold of $\sim 1$~\GeVc\ could bias the ($A$$\times$$\epsilon$) estimation.
To evaluate this effect, the full simulation of the trigger response function is replaced with a parameterization of the trigger response function as a function of transverse momenta.
The parameterization is obtained both in MC and in data by using minimum bias events recorded in parallel with the triggered data sample. 
The analysis is then repeated using either of the two parameterizations, and the resulting difference is taken as the systematic uncertainty.
The effect is small ($< 0.022$) for $p_{\rm T} > 4$~\GeVc, and a maximum uncertainty of 0.070 is estimated for $\lambda_\varphi$ in the first $p_{\rm T}$ interval of the CS frame.
Thirdly, the uncertainty on the detector efficiency includes the uncertainty on the tracking efficiency, the trigger chamber efficiency and the matching between tracks reconstructed in the tracker and in the trigger system.
The resulting uncertainty on the J/$\psi$ yields is evaluated with the same procedure as the one described in~\cite{Abelev:2014qha} and is propagated to the corrected yields of the angular distributions by adding it in quadrature with the statistical ones.
Finally, the fits are redone and the associated uncertainty on $\lambda_\alpha$ parameters is estimated as the square root of the quadratic difference between the new uncertainty returned by the fit and the statistical one.
Its value ranges from 0.046 to 0.133.
This is the main uncertainty for the $\lambda_{\theta\varphi}$ parameter. 

The different sources of systematic uncertainties are summarized in Tab.~\ref{tab:errors}.
The four sources of systematics are independent and can be summed in quadrature to obtain the total systematic uncertainty on each $\lambda_\alpha$ parameter.
Systematic uncertainties are considered uncorrelated among the three polarization parameters and among the $p_{\rm T}$ intervals.

\begin{table*}
\centering 
\begin{tabular}{lllllll}
\hline\noalign{\smallskip}
Source & $\lambda_\theta^{\rm HX}$ & $\lambda_\varphi^{\rm HX}$ & $\lambda_{\theta\varphi}^{\rm HX}$ & $\lambda_\theta^{\rm CS}$ & $\lambda_\varphi^{\rm CS}$ & $\lambda_{\theta\varphi}^{\rm CS}$ \\
\noalign{\smallskip}\hline\noalign{\smallskip}
Signal & 0.035--0.087 & 0.021--0.108 & 0.014--0.032 & 0.022--0.074 & 0.022--0.052 & 0.012--0.063 \\
Inputs & 0.076--0.155 & 0.007--0.024 & 0.006--0.033 & 0.013--0.175 & 0.006--0.040 & 0.004--0.018 \\
Trigger & 0.001--0.064 & 0.001--0.060 & 0.005--0.020 & 0.006--0.036 & 0.007--0.070 & 0.006--0.017 \\
Efficiency & 0.076--0.133 & 0.046--0.069 & 0.064--0.076 & 0.081--0.121 & 0.058--0.072 & 0.073--0.081 \\
\noalign{\smallskip}\hline
\end{tabular}
\caption{Absolute systematic uncertainties on J/$\psi$ polarization parameters in the HX and CS frames. The four different uncertainty sources are the signal extraction (signal), the input distributions of the J/$\psi$ in the MC simulations (inputs), the low-$p_{\rm T}$ trigger response (trigger) and the detector efficiency (efficiency). The last three sources enter in the computation of the acceptance and efficiency factor ($A$$\times$$\epsilon$).}
\label{tab:errors}      
\end{table*}

\section{Results}
\label{sec:results}

\begin{figure}[h]
\centering
\includegraphics[width=1.0\textwidth]{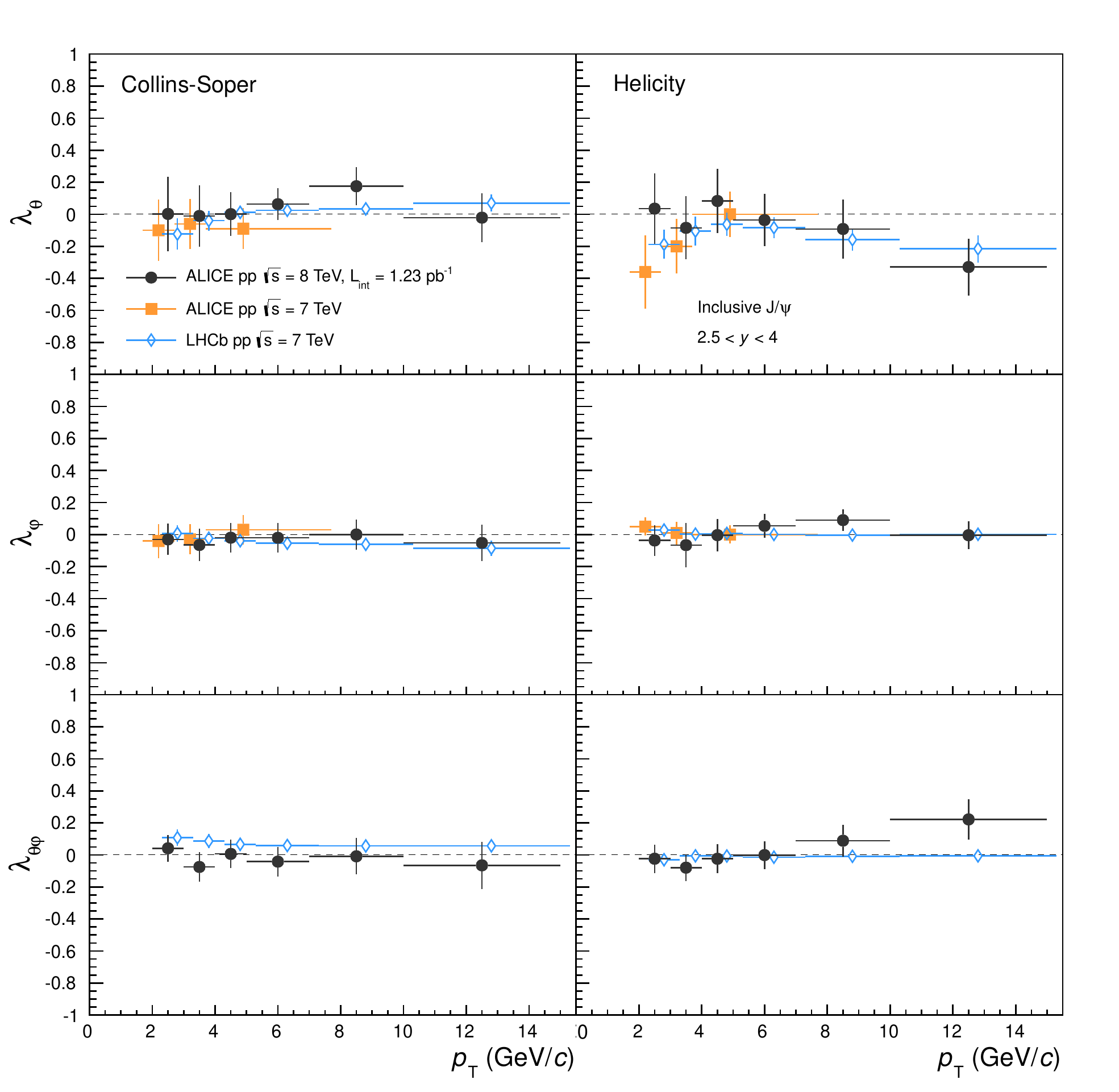}
\caption{(color online) ALICE inclusive J/$\psi$ polarization parameters in pp collisions at $\sqrt{s} = 8$~TeV (black points) compared with ALICE~\cite{Abelev:2011md} inclusive J/$\psi$ (orange squares, shifted horizontally by $- 0.3$~\GeVc) and LHCb~\cite{Aaij:2013nlm} prompt J/$\psi$ (blue open diamonds, shifted horizontally by $+ 0.3$~\GeVc) measurements at $\sqrt{s} = 7$~TeV in the rapidity interval $2.5 < y < 4.0$. The error bars represent the total uncertainties. Left and right plots show results in the Collins-Soper and helicity frames, respectively, for $\lambda_\theta$ (top plots), $\lambda_\varphi$ (middle plots) and $\lambda_{\theta\varphi}$ (bottom plots).}
\label{fig:results_7TeV}   
\end{figure}

The inclusive J/$\psi$ polarization parameters in the interval $2.5 < y < 4.0$ and $2 < p_{\rm T} < 15$~\GeVc\ measured in pp collisions at $\sqrt{s} = 8$~TeV are shown in Fig.~\ref{fig:results_7TeV} for the HX (right) and the CS (left) frames and summarized in Tab.~\ref{tab:resultsHX} and~\ref{tab:resultsCS}, respectively.
In the figure, the error bars represent the total uncertainties computed by adding in quadrature the statistical and systematic uncertainties.
This is the first measurement of the J/$\psi$ polarization parameters at this energy and extends the $p_{\rm T}$ reach of the previous ALICE measurement at $\sqrt{s} = 7$~TeV from 8 to 15~\GeVc. 
The results show that the polarization of inclusive J/$\psi$ mesons is compatible with zero within uncertainties, with a maximum deviation of 1.8 standard deviations away from zero for the highest $p_{\rm T}$ interval for the $\lambda_\theta$ and $\lambda_{\theta\varphi}$ parameters in the HX frame.

\begin{table*}[h]  
\centering     
\begin{tabular}{crrr}
\hline\noalign{\smallskip}
$p_{\rm T}$ (\GeVc) & $\lambda_\theta^{\rm HX}$ & $\lambda_\varphi^{\rm HX}$ & $\lambda_{\theta\varphi}^{\rm HX}$  \\
\noalign{\smallskip}\hline\noalign{\smallskip}
$2-3$ & $0.035 \pm 0.048 \pm 0.215$ & $-0.037 \pm 0.025 \pm 0.093$ & $-0.024 \pm 0.032 \pm 0.082$ \\
$3-4$ & $-0.085 \pm 0.053 \pm 0.189$ & $-0.065 \pm 0.026 \pm 0.134$ & $-0.080 \pm 0.035 \pm 0.077$ \\
$4-5$ & $0.083 \pm 0.066 \pm 0.188$ & $-0.003 \pm 0.033 \pm 0.096$ & $-0.024 \pm 0.043 \pm 0.080$ \\
$5-7$ & $-0.036 \pm 0.058 \pm 0.154$ & $0.055 \pm 0.029 \pm 0.069$ & $-0.001 \pm 0.039 \pm 0.078$ \\
$7-10$ & $-0.092 \pm 0.078 \pm 0.168$ & $0.090 \pm 0.039 \pm 0.056$ & $0.089 \pm 0.055 \pm 0.082$ \\
$10-15$ & $-0.329 \pm 0.121 \pm 0.130$ & $-0.003 \pm 0.070 \pm 0.052$ & $0.222 \pm 0.099 \pm 0.079$ \\
\noalign{\smallskip}\hline
\end{tabular}
\caption{Inclusive J/$\psi$ polarization parameters in the HX frame in the rapidity interval $2.5 < y < 4.0$. The first uncertainty is statistical and the second systematic.}
\label{tab:resultsHX}
\end{table*}

\begin{table*}[h]
\centering     
\begin{tabular}{crrr}
\hline\noalign{\smallskip}
$p_{\rm T}$ (\GeVc) & $\lambda_\theta^{\rm CS}$ & $\lambda_\varphi^{\rm CS}$ & $\lambda_{\theta\varphi}^{\rm CS}$  \\
\noalign{\smallskip}\hline\noalign{\smallskip}
$2-3$ & $0.002 \pm 0.046 \pm 0.228$ & $-0.030 \pm 0.024 \pm 0.095$ & $0.041 \pm 0.032 \pm 0.076$ \\
$3-4$ & $-0.011 \pm 0.052 \pm 0.185$ & $-0.065 \pm 0.026 \pm 0.098$ & $-0.075 \pm 0.035 \pm 0.084$ \\
$4-5$ & $0.001 \pm 0.056 \pm 0.124$ & $-0.019 \pm 0.030 \pm 0.086$ & $0.006 \pm 0.041 \pm 0.080$ \\
$5-7$ & $0.063 \pm 0.048 \pm 0.088$ & $-0.020 \pm 0.031 \pm 0.087$ & $-0.042 \pm 0.041 \pm 0.082$ \\
$7-10$ & $0.175 \pm 0.070 \pm 0.096$ & $0.001 \pm 0.045 \pm 0.082$ & $-0.009 \pm 0.060 \pm 0.096$ \\
$10-15$ & $-0.021 \pm 0.110 \pm 0.106$ & $-0.052 \pm 0.084 \pm 0.077$ & $-0.065 \pm 0.110 \pm 0.098$ \\
\noalign{\smallskip}\hline
\end{tabular}
\caption{Inclusive J/$\psi$ polarization parameters in the CS frame in the rapidity interval $2.5 < y < 4.0$. The first uncertainty is statistical and the second systematic.}
\label{tab:resultsCS}  
\end{table*}

As the differences between the J/$\psi$ polarization in pp collisions at $\sqrt{s} = 7$~TeV and 8~TeV are expected to be negligible (see Kniehl et al. predictions in Ref.~\cite{Aaij:2013nlm} and in this paper), the measurements at the two energies can be directly compared.
This comparison is shown in Fig.~\ref{fig:results_7TeV} with the published results by ALICE~\cite{Abelev:2011md} (inclusive J/$\psi$) and LHCb~\cite{Aaij:2013nlm} (prompt J/$\psi$, i.e.\ without the contribution from b-hadron decays) in the same rapidity interval for pp collisions at $\sqrt{s} = 7$~TeV.   
The two ALICE measurements agree within one standard deviation. 
Concerning the comparison between ALICE and LHCb results, a rather good agreement is observed for all polarization parameters over the full $p_{\rm T}$ interval. 
The observed agreement between the ALICE and LHCb results seems to indicate that J/$\psi$ from b-hadron decays do not introduce any observable difference in the polarization parameters.

\begin{figure}[h]
\centering
\includegraphics[width=1.0\textwidth]{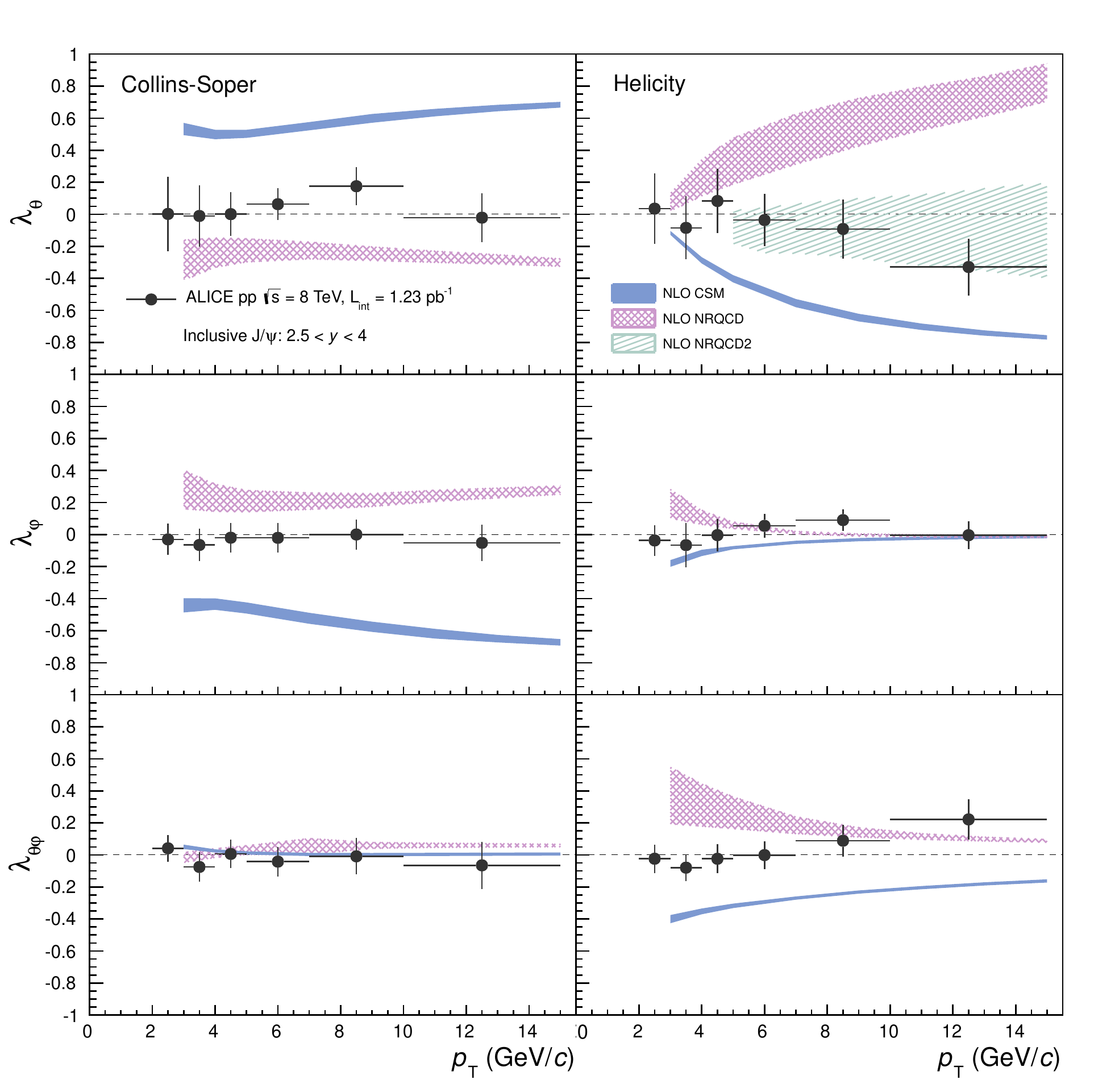}
\caption{(color online) Inclusive J/$\psi$ polarization parameters in pp collisions at $\sqrt{s} = 8$~TeV (black points, error bars represent the total uncertainties) compared with model predictions: NLO CSM~\cite{butenschoen:2012px} (blue filled bands), NRQCD~\cite{butenschoen:2012px} (red shaded bands) and NRQCD2~\cite{Chao:2012iv} (light blue hatched band). Left and right plots show the results in the Collins-Soper and helicity frames, respectively, for $\lambda_\theta$ (top plots), $\lambda_\varphi$ (middle plots) and $\lambda_{\theta\varphi}$ (bottom plots).}
\label{fig:results_models}   
\end{figure}

Figure~\ref{fig:results_models} shows the comparison of all the measured polarization parameters with the NLO CSM (blue filled band) and NRQCD (red shaded band) predictions from~\cite{butenschoen:2012px} and with another NRQCD (light blue hatched band) prediction from~\cite{Chao:2012iv} for $\lambda_\theta$ in the helicity frame (labeled as NLO NRQCD2 in Fig.~\ref{fig:results_models}). 
The shown error bands of the models are evaluated by adding in quadrature the uncertainties due to the different scale variations (renormalization, factorization and NRQCD scales) in the calculation and LDME variations.
The difference between the two NRQCD calculations originates from the data used to compute the LDMEs.
Moreover, in~\cite{butenschoen:2012px} only direct J/$\psi$ (i.e.\ without feed-down from excited states) are considered, while in~\cite{Chao:2012iv} feed-down from excited states is included in the J/$\psi$ prediction.

The CSM and NRQCD calculations from~\cite{butenschoen:2012px} predict an opposite $p_{\rm T}$ trend for all polarization parameters in the two frames.
The $p_{\rm T}$ dependence is relatively small over the considered $p_{\rm T}$ interval, except for the $\lambda_\theta$ parameter in the HX frame.
The NRQCD calculation including both color-singlet and color-octet contributions provides a qualitatively better description of the J/$\psi$ polarization measurement, except for $\lambda_\theta$ in the HX frame where the large transverse J/$\psi$ polarization predicted by the NRQCD~\cite{butenschoen:2012px} is in contradiction with the experimental observations. 
The NRQCD prediction from~\cite{Chao:2012iv} favours either zero or small longitudinal polarization, with large theoretical uncertainties, and shows a good agreement with the measurements in the intermediate $p_{\rm T}$ interval ($5 < p_{\rm T} < 15$~\GeVc), but gives no prediction for $p_{\rm T} < 5$~\GeVc.
This agreement is not surprising because this model includes the measurements of the J/$\psi$ polarization performed at Tevatron~\cite{Affolder:2000nn,Abulencia:2007us} to determine the LDMEs. 
As this model gives no prediction for the other polarization parameters in the HX frame, as well as for the whole set of polarization parameters in the CS frame, it is difficult to draw a clear conclusion about its ability to describe the measurements.  

\begin{figure}[h]
\centering
\includegraphics[width=0.70\textwidth]{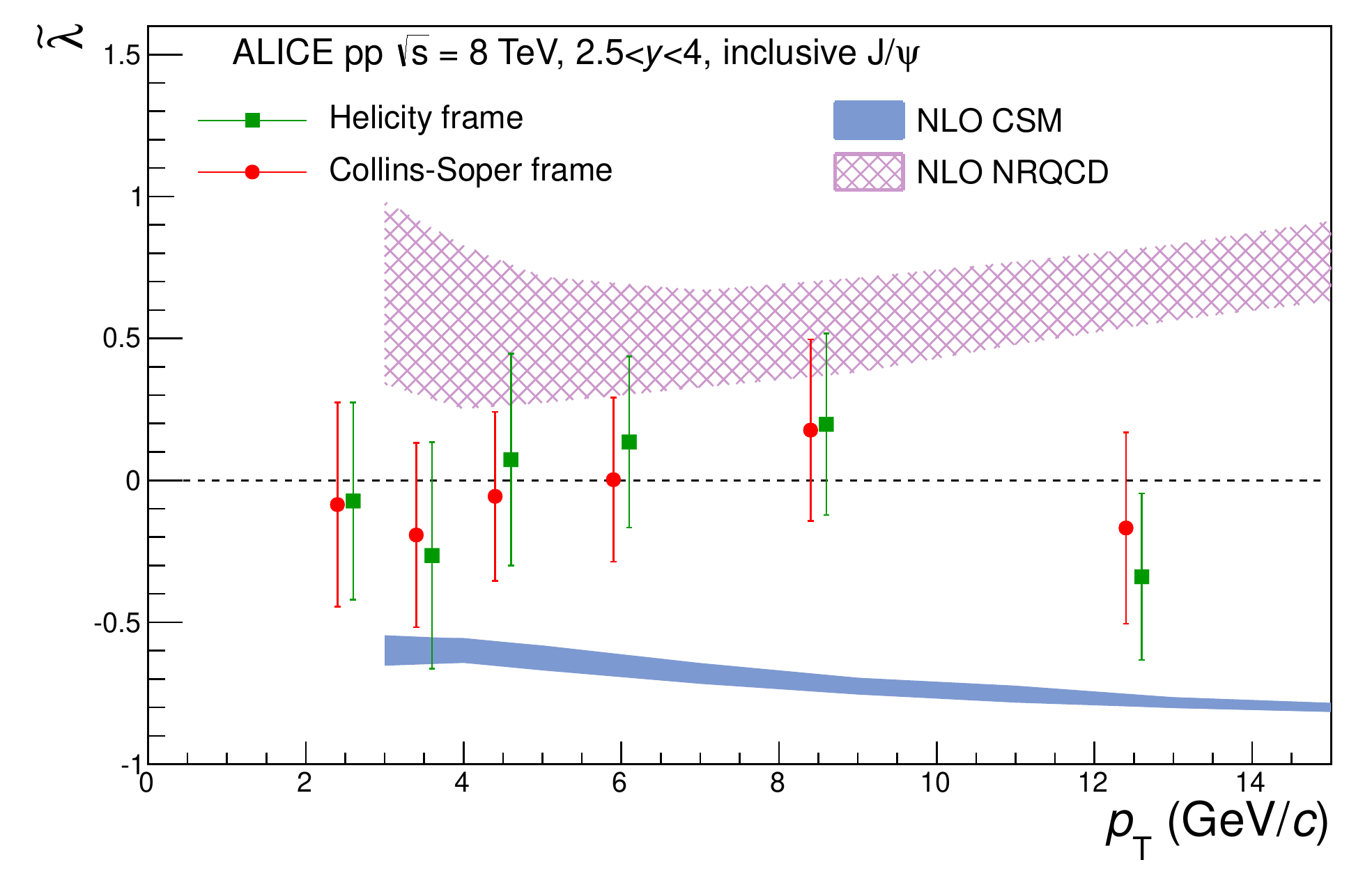}
\caption{(color online) Inclusive J/$\psi$ frame-invariant quantity $\widetilde{\lambda}$ in pp collisions at $\sqrt{s} = 8$~TeV in the Collins-Soper (red points, shifted horizontally by $-0.1$~\GeVc) and helicity (green squares, shifted horizontally by $+0.1$~\GeVc) frames compared with the NLO CSM (blue full band) and NRQCD (red shaded band) model predictions from~\cite{butenschoen:2012px}.}
\label{fig:invariant}   
\end{figure}

As shown by Faccioli et al.~\cite{Faccioli:2010kd}, frame-invariant observables do exist and the most commonly considered one is  
\begin{equation}
\widetilde{\lambda} = \frac{\lambda_\theta + 3 \lambda_\varphi}{1 - \lambda_\varphi} .
\end{equation} 
Figure~\ref{fig:invariant} shows the $p_{\rm T}$ dependence of this invariant quantity for both frames in comparison with the NLO CSM and NRQCD predictions from~\cite{butenschoen:2012px}.
To propagate the uncertainties on $\lambda_\theta$ and $\lambda_\varphi$ to the frame-invariant quantity $\widetilde{\lambda}$, the correlation coefficient $\rho_{\lambda_\theta,\lambda_\varphi}$ returned by the simultaneous fit of the angular distributions are taken into account to compute the statistical uncertainties, while the systematic uncertainties are assumed to be uncorrelated.
For the model predictions, the quoted error bands are computed by adding the uncertainties due to the different scales and LDME variations in quadrature, after propagation of the correlated effects between $\lambda_\theta$ and $\lambda_\varphi$.
The comparison of the frame-invariant quantity $\widetilde{\lambda}$ shows that the ALICE measurements in both frames are in good agreement within uncertainties, confirming the consistency of the results. 
Both the CSM and the NRQCD model respect the frame invariance for $\widetilde{\lambda}$, but clearly none of them is able to describe the measured $p_{\rm T}$ dependence, even if the NRQCD prediction shows a better agreement with data ($\chi^2_{\rm /NDF} = 1.7$ compared to $\chi^2_{\rm /NDF} = 2.0$ by CSM), although with large uncertainties especially for $p_{\rm T} < 6$~\GeVc.

Using the ALICE inclusive J/$\psi$ cross section measurement at $\sqrt{s} = 8$~TeV~\cite{Adam:2015rta}, an average value for the polarization parameters over $p_{\rm T}$ can be computed in the following way 
\begin{equation}
\langle\lambda_\alpha\rangle = \frac{1}{\sigma_{\rm tot}} \sum_{j=1}^6 \sigma_j \lambda_\alpha^j ,
\end{equation}
with
\begin{equation}
\sigma_{\rm tot} = \sum_{j=1}^6 \sigma_j .
\end{equation}
In these equations, $j$ is running over the six $p_{\rm T}$ bins of this analysis, $\sigma_j$ is the integrated inclusive J/$\psi$ cross section in the $p_{\rm T}$ bin $j$  and $\lambda_\alpha^j$ is the measured polarization parameter in the corresponding bin.
The resulting average values of the polarization parameters over $2 < p_{\rm T} < 15$~\GeVc\ are summarized in Tab.~\ref{tab:average}.
The uncertainties are computed by propagating the total uncertainty on the polarization parameters and the uncorrelated uncertainty on the cross section measurements from~\cite{Adam:2015rta}.
All averaged values of the polarization parameters are consistent with zero within uncertainties. 

\begin{figure}[h]
\centering
\includegraphics[width=0.43\textwidth]{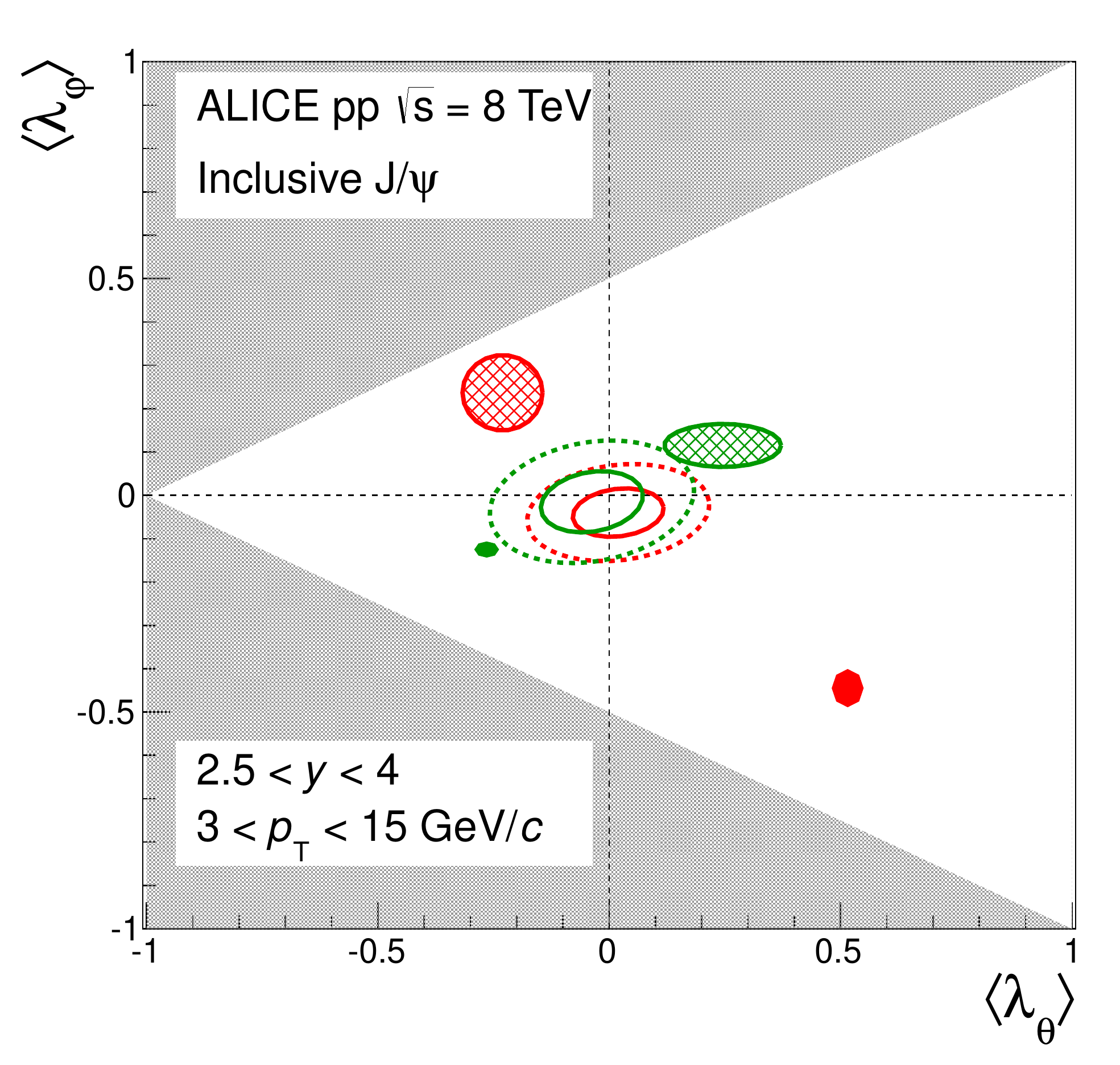} \hfill
\includegraphics[width=0.43\textwidth]{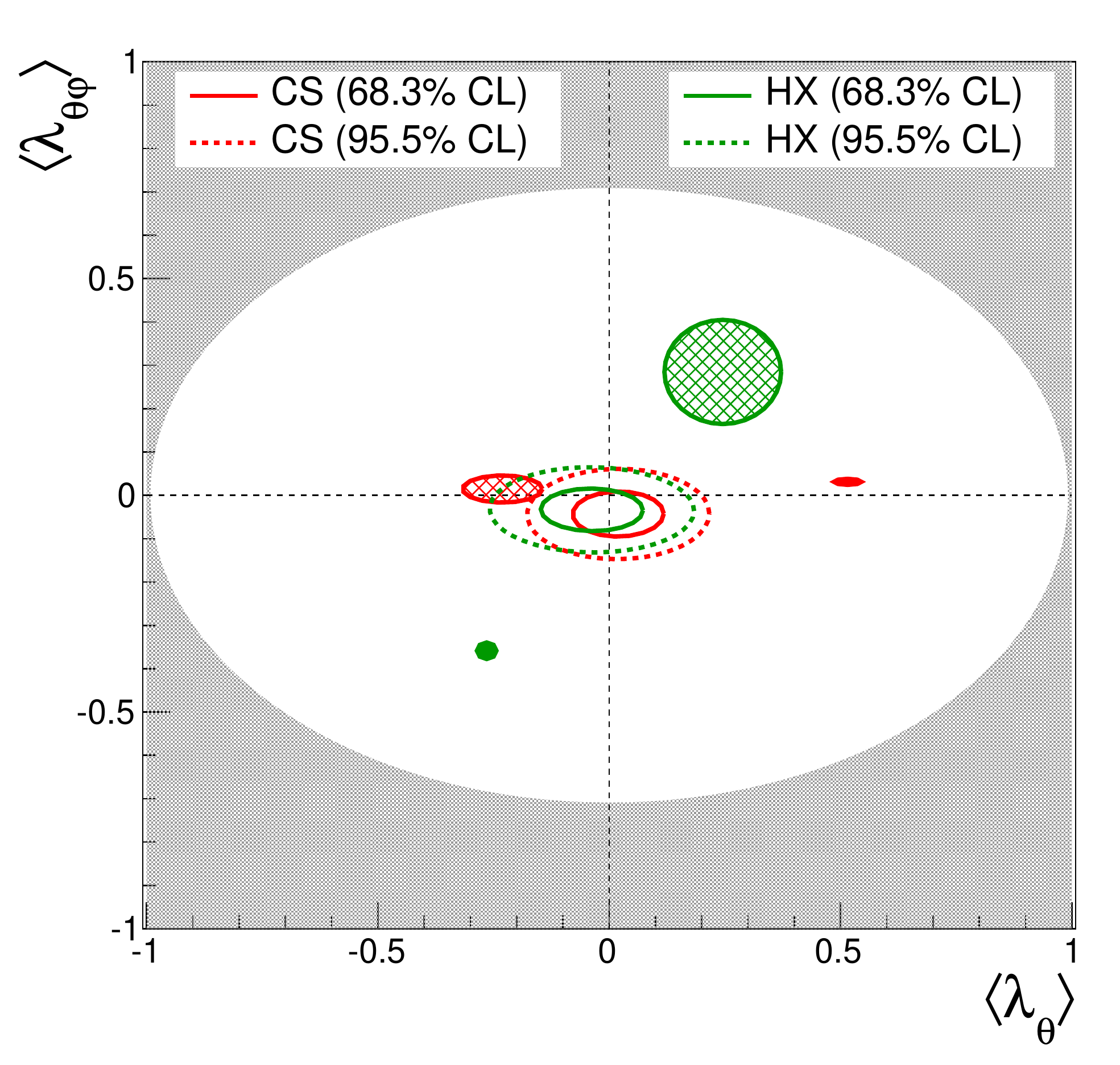} \\
\includegraphics[width=0.43\textwidth]{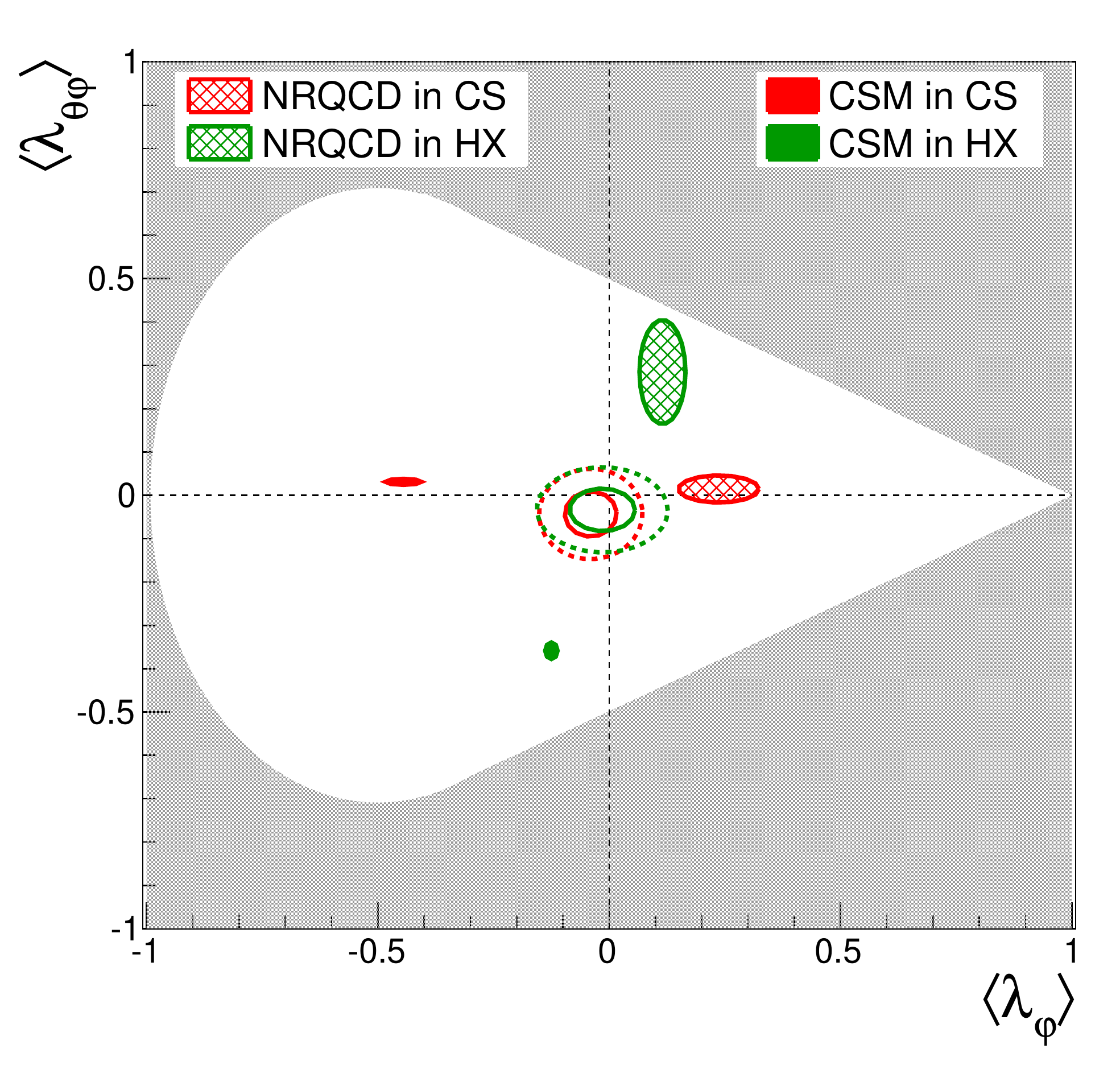}
\caption{(color online) Average $p_{\rm T}$-integrated (in rapidity range $2.5 < y < 4.0$) inclusive J/$\psi$ polarization parameters $\langle\lambda_\theta\rangle$, $\langle\lambda_\varphi\rangle$ and $\langle\lambda_{\theta\varphi}\rangle$ in allowed 2-D regions (white areas) for $3 < p_{\rm T} < 15$~\GeVc. Full (dashed) ellipses show 1-$\sigma$ (2-$\sigma$) contours in Collins-Sopper (CS, red) and helicity (HX, green) frames. Model predictions~\cite{butenschoen:2012px} are represented by filled contours, full filled for the CSM and shaded filled for the NRQCD model, in green for the HX frame and in red for the CS frame.}
\label{fig:2D}
\end{figure}

The $p_{\rm T}$-integrated values can be used to check the consistency of the measured polarization parameters with respect to the theoretically allowed parameter space in 2-D plots, as shown in Fig.~\ref{fig:2D} for $3 < p_{\rm T} < 15$~\GeVc.
This figure takes into account the correlation coefficients $\rho_{\lambda_\theta,\lambda_\varphi}$, $\rho_{\lambda_\theta,\lambda_{\theta\varphi}}$ and $\rho_{\lambda_\varphi,\lambda_{\theta\varphi}}$ between the polarization parameters returned by the simultaneous fit of the angular distributions.
Their values are averaged over $p_{\rm T}$ as for the $\lambda_\alpha$.
The average coefficient correlations, in both HX and CS frames, are in the range $[-0.05\,;\,0.05]$ for $\rho_{\lambda_\theta,\lambda_{\theta\varphi}}$ and $\rho_{\lambda_\varphi,\lambda_{\theta\varphi}}$, while $\rho_{\lambda_\theta,\lambda_\varphi}$ is about 0.2.  
Contour ellipses show that the $p_{\rm T}$-integrated polarization parameters are well within the allowed theoretical parameter-space and highlight the observed absence of polarization of inclusive J/$\psi$ at forward rapidity in pp collisions at $\sqrt{s} = 8$~TeV.
The comparison with the $p_{\rm T}$-integrated NLO CSM and NRQCD predictions is shown in Fig.~\ref{fig:2D} (right). 
These 2-D plots confirm the difficulty of the models to reproduce the ALICE measurements and show also that the discrepancy from data is larger for the CSM than for the NRQCD calculation, especially in the plane $(\lambda_\theta,\lambda_\varphi)$ in the CS frame.

\begin{table*}
\centering   
\begin{tabular}{ccc}
\hline\noalign{\smallskip}
Parameter & HX frame & CS frame  \\
\noalign{\smallskip}\hline\noalign{\smallskip}
$\langle \lambda_\theta \rangle$ & $-0.006 \pm 0.115$ & $0.012 \pm 0.116$ \\
$\langle \lambda_{\varphi} \rangle$ & $-0.024 \pm 0.058$ & $-0.036 \pm 0.053$ \\
$\langle \lambda_{\theta\varphi} \rangle$ & $-0.029 \pm 0.047$ & $-0.006 \pm 0.047$ \\
\noalign{\smallskip}\hline
\end{tabular}
\caption{Average $p_{\rm T}$-integrated (over $2 < p_{\rm T} < 15$~\GeVc\ in the rapidity range $2.5 < y < 4.0$) inclusive J/$\psi$ polarization parameters $\langle\lambda_\theta\rangle$, $\langle\lambda_\varphi\rangle$ and $\langle\lambda_{\theta\varphi}\rangle$ in the HX and CS frames.}
\label{tab:average}   
\end{table*}

\section{Conclusion}
\label{sec:conclusion}

The polarization parameters of inclusive J/$\psi$ mesons are measured with the ALICE detector at forward rapidity ($2.5 < y < 4.0$) in pp collisions at $\sqrt{s} = 8$~TeV.
Detailed investigations of their transverse momentum dependence in the interval $2 < p_{\rm T} < 15$~\GeVc\ show that no polarization is observed for the measured J/$\psi$ mesons.
This result is further highlighted by the $p_{\rm T}$-integrated polarization parameters.
The comparisons with the theoretical predictions from the Color-Singlet Model and the Non-Relativistic QCD model show that none of the two approaches is able to describe all polarization parameters over the studied $p_{\rm T}$ interval.
It follows that a full understanding of the production mechanism of J/$\psi$ in hadronic collisions remains an open question.

%
%

\newenvironment{acknowledgement}{\relax}{\relax}
\begin{acknowledgement}
\section*{Acknowledgements}

The ALICE Collaboration would like to thank all its engineers and technicians for their invaluable contributions to the construction of the experiment and the CERN accelerator teams for the outstanding performance of the LHC complex.
The ALICE Collaboration gratefully acknowledges the resources and support provided by all Grid centres and the Worldwide LHC Computing Grid (WLCG) collaboration.
The ALICE Collaboration acknowledges the following funding agencies for their support in building and running the ALICE detector:
A. I. Alikhanyan National Science Laboratory (Yerevan Physics Institute) Foundation (ANSL), State Committee of Science and World Federation of Scientists (WFS), Armenia;
Austrian Academy of Sciences and Nationalstiftung f\"{u}r Forschung, Technologie und Entwicklung, Austria;
Ministry of Communications and High Technologies, National Nuclear Research Center, Azerbaijan;
Conselho Nacional de Desenvolvimento Cient\'{\i}fico e Tecnol\'{o}gico (CNPq), Universidade Federal do Rio Grande do Sul (UFRGS), Financiadora de Estudos e Projetos (Finep) and Funda\c{c}\~{a}o de Amparo \`{a} Pesquisa do Estado de S\~{a}o Paulo (FAPESP), Brazil;
Ministry of Science \& Technology of China (MSTC), National Natural Science Foundation of China (NSFC) and Ministry of Education of China (MOEC) , China;
Ministry of Science and Education, Croatia;
Ministry of Education, Youth and Sports of the Czech Republic, Czech Republic;
The Danish Council for Independent Research | Natural Sciences, the Carlsberg Foundation and Danish National Research Foundation (DNRF), Denmark;
Helsinki Institute of Physics (HIP), Finland;
Commissariat \`{a} l'Energie Atomique (CEA) and Institut National de Physique Nucl\'{e}aire et de Physique des Particules (IN2P3) and Centre National de la Recherche Scientifique (CNRS), France;
Bundesministerium f\"{u}r Bildung, Wissenschaft, Forschung und Technologie (BMBF) and GSI Helmholtzzentrum f\"{u}r Schwerionenforschung GmbH, Germany;
General Secretariat for Research and Technology, Ministry of Education, Research and Religions, Greece;
National Research, Development and Innovation Office, Hungary;
Department of Atomic Energy Government of India (DAE), Department of Science and Technology, Government of India (DST), University Grants Commission, Government of India (UGC) and Council of Scientific and Industrial Research (CSIR), India;
Indonesian Institute of Science, Indonesia;
Centro Fermi - Museo Storico della Fisica e Centro Studi e Ricerche Enrico Fermi and Istituto Nazionale di Fisica Nucleare (INFN), Italy;
Institute for Innovative Science and Technology , Nagasaki Institute of Applied Science (IIST), Japan Society for the Promotion of Science (JSPS) KAKENHI and Japanese Ministry of Education, Culture, Sports, Science and Technology (MEXT), Japan;
Consejo Nacional de Ciencia (CONACYT) y Tecnolog\'{i}a, through Fondo de Cooperaci\'{o}n Internacional en Ciencia y Tecnolog\'{i}a (FONCICYT) and Direcci\'{o}n General de Asuntos del Personal Academico (DGAPA), Mexico;
Nederlandse Organisatie voor Wetenschappelijk Onderzoek (NWO), Netherlands;
The Research Council of Norway, Norway;
Commission on Science and Technology for Sustainable Development in the South (COMSATS), Pakistan;
Pontificia Universidad Cat\'{o}lica del Per\'{u}, Peru;
Ministry of Science and Higher Education and National Science Centre, Poland;
Korea Institute of Science and Technology Information and National Research Foundation of Korea (NRF), Republic of Korea;
Ministry of Education and Scientific Research, Institute of Atomic Physics and Romanian National Agency for Science, Technology and Innovation, Romania;
Joint Institute for Nuclear Research (JINR), Ministry of Education and Science of the Russian Federation and National Research Centre Kurchatov Institute, Russia;
Ministry of Education, Science, Research and Sport of the Slovak Republic, Slovakia;
National Research Foundation of South Africa, South Africa;
Centro de Aplicaciones Tecnol\'{o}gicas y Desarrollo Nuclear (CEADEN), Cubaenerg\'{\i}a, Cuba and Centro de Investigaciones Energ\'{e}ticas, Medioambientales y Tecnol\'{o}gicas (CIEMAT), Spain;
Swedish Research Council (VR) and Knut \& Alice Wallenberg Foundation (KAW), Sweden;
European Organization for Nuclear Research, Switzerland;
National Science and Technology Development Agency (NSDTA), Suranaree University of Technology (SUT) and Office of the Higher Education Commission under NRU project of Thailand, Thailand;
Turkish Atomic Energy Agency (TAEK), Turkey;
National Academy of  Sciences of Ukraine, Ukraine;
Science and Technology Facilities Council (STFC), United Kingdom;
National Science Foundation of the United States of America (NSF) and United States Department of Energy, Office of Nuclear Physics (DOE NP), United States of America.
\end{acknowledgement}

\bibliographystyle{utphys}   
\bibliography{alicepreprint_CDS-v8}

\providecommand{\href}[2]{#2}\begingroup\raggedright\begin{thebibliography}{10}

\bibitem{brambilla:2010cs}
N.~Brambilla, S.~Eidelman, B.~Heltsley, R.~Vogt, G.~Bodwin, {\em et~al.},
  ``{Heavy quarkonium: progress, puzzles, and opportunities},''
  \href{http://dx.doi.org/10.1140/epjc/s10052-010-1534-9}{{\em Eur. Phys. J.}
  {\bfseries C71} (2011) 1534},
\href{http://arxiv.org/abs/1010.5827}{{\ttfamily arXiv:1010.5827 [hep-ph]}}.

\bibitem{Andronic:2015wma}
A.~Andronic {\em et~al.}, ``{Heavy-flavour and quarkonium production in the LHC
  era: from proton-proton to heavy-ion collisions},''
  \href{http://dx.doi.org/10.1140/epjc/s10052-015-3819-5}{{\em Eur. Phys. J.}
  {\bfseries C76} (2016) 107},
\href{http://arxiv.org/abs/1506.03981}{{\ttfamily arXiv:1506.03981 [nucl-ex]}}.

\bibitem{fritzsch:1977ay}
H.~Fritzsch, ``{Producing Heavy Quark Flavors in Hadronic Collisions: A Test of
  Quantum Chromodynamics},''
\href{http://dx.doi.org/10.1016/0370-2693(77)90108-3}{{\em Phys. Lett.}
  {\bfseries B67} (1977) 217}.

\bibitem{Bodwin:1994jh}
G.~T. Bodwin, E.~Braaten, and G.~P. Lepage, ``{Rigorous QCD analysis of
  inclusive annihilation and production of heavy quarkonium},''
  \href{http://dx.doi.org/10.1103/PhysRevD.55.5853,
  10.1103/PhysRevD.51.1125}{{\em Phys. Rev.} {\bfseries D51} (1995)
  1125--1171}, \href{http://arxiv.org/abs/hep-ph/9407339}{{\ttfamily
  arXiv:hep-ph/9407339 [hep-ph]}}.
[Erratum: Phys. Rev.D55,5853(1997)].

\bibitem{Einhorn:1975ua}
M.~B. Einhorn and S.~D. Ellis, ``{Hadronic Production of the New Resonances:
  Probing Gluon Distributions},''
\href{http://dx.doi.org/10.1103/PhysRevD.12.2007}{{\em Phys. Rev.} {\bfseries
  D12} (1975) 2007}.

\bibitem{Lansberg:2008gk}
J.~P. Lansberg, ``{On the mechanisms of heavy-quarkonium hadroproduction},''
  \href{http://dx.doi.org/10.1140/epjc/s10052-008-0826-9}{{\em Eur. Phys. J.}
  {\bfseries C61} (2009) 693--703},
\href{http://arxiv.org/abs/0811.4005}{{\ttfamily arXiv:0811.4005 [hep-ph]}}.

\bibitem{Acharya:2017hjh}
{\bfseries ALICE} Collaboration, S.~Acharya {\em et~al.}, ``{Energy dependence
  of forward-rapidity $\mathrm {J}/\psi $ and $\psi \mathrm {(2S)}$ production
  in pp collisions at the LHC},''
  \href{http://dx.doi.org/10.1140/epjc/s10052-017-4940-4}{{\em Eur. Phys. J.}
  {\bfseries C77} (2017) 392},
\href{http://arxiv.org/abs/1702.00557}{{\ttfamily arXiv:1702.00557 [hep-ex]}}.

\bibitem{Aaij:2014bga}
{\bfseries LHCb} Collaboration, R.~Aaij {\em et~al.}, ``{Measurement of the
  $\eta_c (1S)$ production cross-section in proton-proton collisions via the
  decay $\eta_c (1S) \rightarrow p \bar{p}$},''
  \href{http://dx.doi.org/10.1140/epjc/s10052-015-3502-x}{{\em Eur. Phys. J.}
  {\bfseries C75} no.~7, (2015) 311},
\href{http://arxiv.org/abs/1409.3612}{{\ttfamily arXiv:1409.3612 [hep-ex]}}.

\bibitem{Butenschoen:2014dra}
M.~Butenschoen, Z.-G. He, and B.~A. Kniehl, ``{$\eta_c$ production at the LHC
  challenges nonrelativistic-QCD factorization},''
  \href{http://dx.doi.org/10.1103/PhysRevLett.114.092004}{{\em Phys. Rev.
  Lett.} {\bfseries 114} no.~9, (2015) 092004},
\href{http://arxiv.org/abs/1411.5287}{{\ttfamily arXiv:1411.5287 [hep-ph]}}.

\bibitem{butenschoen:2012px}
M.~Butenschoen and B.~A. Kniehl, ``{J/$\psi$ polarization at Tevatron and LHC:
  Nonrelativistic-QCD factorization at the crossroads},''
  \href{http://dx.doi.org/10.1103/PhysRevLett.108.172002}{{\em Phys. Rev.
  Lett.} {\bfseries 108} (2012) 172002},
\href{http://arxiv.org/abs/1201.1872}{{\ttfamily arXiv:1201.1872 [hep-ph]}}.

\bibitem{Faccioli:2010kd}
P.~Faccioli, C.~Lourenco, J.~Seixas, and H.~K. Woehri, ``{Towards the
  experimental clarification of quarkonium polarization},''
  \href{http://dx.doi.org/10.1140/epjc/s10052-010-1420-5}{{\em Eur. Phys. J.}
  {\bfseries C69} (2010) 657--673},
\href{http://arxiv.org/abs/1006.2738}{{\ttfamily arXiv:1006.2738 [hep-ph]}}.

\bibitem{Chatrchyan:2013cla}
{\bfseries CMS} Collaboration, S.~Chatrchyan {\em et~al.}, ``{Measurement of
  the prompt $J/\psi$ and $\psi$(2S) polarizations in $pp$ collisions at
  $\sqrt{s}$ = 7 TeV},''
  \href{http://dx.doi.org/10.1016/j.physletb.2013.10.055}{{\em Phys. Lett.}
  {\bfseries B727} (2013) 381--402},
\href{http://arxiv.org/abs/1307.6070}{{\ttfamily arXiv:1307.6070 [hep-ex]}}.

\bibitem{Abelev:2011md}
{\bfseries ALICE} Collaboration, B.~Abelev {\em et~al.}, ``{$J/\psi$
  polarization in $pp$ collisions at $\sqrt{s}=7$ TeV},''
  \href{http://dx.doi.org/10.1103/PhysRevLett.108.082001}{{\em Phys. Rev.
  Lett.} {\bfseries 108} (2012) 082001},
\href{http://arxiv.org/abs/1111.1630}{{\ttfamily arXiv:1111.1630 [hep-ex]}}.

\bibitem{Aaij:2013nlm}
{\bfseries LHCb} Collaboration, R.~Aaij {\em et~al.}, ``{Measurement of
  $J/\psi$ polarization in $pp$ collisions at $\sqrt{s}=7$ TeV},''
  \href{http://dx.doi.org/10.1140/epjc/s10052-013-2631-3}{{\em Eur. Phys. J.}
  {\bfseries C73} no.~11, (2013) 2631},
\href{http://arxiv.org/abs/1307.6379}{{\ttfamily arXiv:1307.6379 [hep-ex]}}.

\bibitem{Aamodt:2008zz}
{\bfseries ALICE} Collaboration, K.~Aamodt {\em et~al.}, ``{The ALICE
  experiment at the CERN LHC},''
\href{http://dx.doi.org/10.1088/1748-0221/3/08/S08002}{{\em JINST} {\bfseries
  3} (2008) S08002}.

\bibitem{Abelev:2014ffa}
{\bfseries ALICE} Collaboration, B.~B. Abelev {\em et~al.}, ``{Performance of
  the ALICE Experiment at the CERN LHC},''
  \href{http://dx.doi.org/10.1142/S0217751X14300440}{{\em Int. J. Mod. Phys.}
  {\bfseries A29} (2014) 1430044},
\href{http://arxiv.org/abs/1402.4476}{{\ttfamily arXiv:1402.4476 [nucl-ex]}}.

\bibitem{Aamodt:2011gj}
{\bfseries ALICE} Collaboration, K.~Aamodt {\em et~al.}, ``{Rapidity and
  transverse momentum dependence of inclusive J$/\psi$ production in $pp$
  collisions at $\sqrt{s} = 7$ TeV},''
  \href{http://dx.doi.org/10.1016/j.physletb.2011.09.054,
  10.1016/j.physletb.2012.10.060}{{\em Phys. Lett.} {\bfseries B704} (2011)
  442--455}, \href{http://arxiv.org/abs/1105.0380}{{\ttfamily arXiv:1105.0380
  [hep-ex]}}.
[Erratum: Phys. Lett.B718,692(2012)].

\bibitem{Aamodt:2010aa}
{\bfseries ALICE} Collaboration, K.~Aamodt {\em et~al.}, ``{Alignment of the
  ALICE Inner Tracking System with cosmic-ray tracks},''
  \href{http://dx.doi.org/10.1088/1748-0221/5/03/P03003}{{\em JINST} {\bfseries
  5} (2010) P03003},
\href{http://arxiv.org/abs/1001.0502}{{\ttfamily arXiv:1001.0502
  [physics.ins-det]}}.

\bibitem{Adam:2015rta}
{\bfseries ALICE} Collaboration, J.~Adam {\em et~al.}, ``{Inclusive quarkonium
  production at forward rapidity in pp collisions at $\sqrt{s}=8$ TeV},''
  \href{http://dx.doi.org/10.1140/epjc/s10052-016-3987-y}{{\em Eur. Phys. J.}
  {\bfseries C76} no.~4, (2016) 184},
\href{http://arxiv.org/abs/1509.08258}{{\ttfamily arXiv:1509.08258 [hep-ex]}}.

\bibitem{ALICE-PUBLIC-2015-006}
{\bfseries ALICE} Collaboration, ``{Quarkonium signal extraction in ALICE},''.
  \url{https://cds.cern.ch/record/2060096}.

\bibitem{Lange:2001uf}
D.~J. Lange, ``{The EvtGen particle decay simulation package},''
\href{http://dx.doi.org/10.1016/S0168-9002(01)00089-4}{{\em Nucl. Instrum.
  Meth.} {\bfseries A462} (2001) 152--155}.

\bibitem{Barberio:1990ms}
E.~Barberio, B.~van Eijk, and Z.~Was, ``{PHOTOS: A Universal Monte Carlo for
  QED radiative corrections in decays},''
\href{http://dx.doi.org/10.1016/0010-4655(91)90012-A}{{\em Comput. Phys.
  Commun.} {\bfseries 66} (1991) 115--128}.

\bibitem{Spiridonov:2004mp}
A.~Spiridonov, ``{Bremsstrahlung in leptonic onia decays: Effects on mass
  spectra},''
\href{http://arxiv.org/abs/hep-ex/0510076}{{\ttfamily arXiv:hep-ex/0510076
  [hep-ex]}}.

\bibitem{Brun:1994aa}
R.~Brun, F.~Bruyant, F.~Carminati, S.~Giani, M.~Maire, A.~McPherson,
  G.~Patrick, and L.~Urban, ``{GEANT Detector Description and Simulation
  Tool},''
\href{http://dx.doi.org/10.17181/CERN.MUHF.DMJ1}{{\em CERN-W-5013} (1994) }.

\bibitem{Abelev:2014qha}
{\bfseries ALICE} Collaboration, B.~B. Abelev {\em et~al.}, ``{Measurement of
  quarkonium production at forward rapidity in $pp$ collisions at $\sqrt{s} =
  7$ TeV},'' \href{http://dx.doi.org/10.1140/epjc/s10052-014-2974-4}{{\em Eur.
  Phys. J.} {\bfseries C74} no.~8, (2014) 2974},
\href{http://arxiv.org/abs/1403.3648}{{\ttfamily arXiv:1403.3648 [nucl-ex]}}.

\bibitem{Chao:2012iv}
K.-T. Chao, Y.-Q. Ma, H.-S. Shao, K.~Wang, and Y.-J. Zhang, ``{$J/\psi$
  Polarization at Hadron Colliders in Nonrelativistic QCD},''
  \href{http://dx.doi.org/10.1103/PhysRevLett.108.242004}{{\em Phys. Rev.
  Lett.} {\bfseries 108} (2012) 242004},
\href{http://arxiv.org/abs/1201.2675}{{\ttfamily arXiv:1201.2675 [hep-ph]}}.

\bibitem{Affolder:2000nn}
{\bfseries CDF} Collaboration, T.~Affolder {\em et~al.}, ``{Measurement of
  $J/\psi$ and $\psi(2S)$ polarization in $p\bar{p}$ collisions at $\sqrt{s} =
  1.8$ TeV},'' \href{http://dx.doi.org/10.1103/PhysRevLett.85.2886}{{\em Phys.
  Rev. Lett.} {\bfseries 85} (2000) 2886--2891},
\href{http://arxiv.org/abs/hep-ex/0004027}{{\ttfamily arXiv:hep-ex/0004027
  [hep-ex]}}.

\bibitem{Abulencia:2007us}
{\bfseries CDF} Collaboration, A.~Abulencia {\em et~al.}, ``{Polarization of
  $J/\psi$ and $\psi_{2S}$ mesons produced in $p \bar{p}$ collisions at
  $\sqrt{s}$ = 1.96-TeV},''
  \href{http://dx.doi.org/10.1103/PhysRevLett.99.132001}{{\em Phys. Rev. Lett.}
  {\bfseries 99} (2007) 132001},
\href{http://arxiv.org/abs/0704.0638}{{\ttfamily arXiv:0704.0638 [hep-ex]}}.

\end{thebibliography}\endgroup

\newpage
\appendix
\section{The ALICE Collaboration}
\label{app:collab}

\begingroup
\small
\begin{flushleft}
S.~Acharya\Irefn{org138}\And 
F.T.-.~Acosta\Irefn{org22}\And 
D.~Adamov\'{a}\Irefn{org94}\And 
J.~Adolfsson\Irefn{org81}\And 
M.M.~Aggarwal\Irefn{org98}\And 
G.~Aglieri Rinella\Irefn{org36}\And 
M.~Agnello\Irefn{org33}\And 
N.~Agrawal\Irefn{org49}\And 
Z.~Ahammed\Irefn{org138}\And 
S.U.~Ahn\Irefn{org77}\And 
S.~Aiola\Irefn{org143}
% ...
\And
\renewcommand\labelenumi{\textsuperscript{\theenumi}~}

\section*{Affiliation notes}
\renewcommand\theenumi{\roman{enumi}}
\begin{Authlist}
\item \Adef{org*}Deceased
\item \Adef{orgI}Dipartimento DET del Politecnico di Torino, Turin, Italy
\item \Adef{orgII}M.V. Lomonosov Moscow State University, D.V. Skobeltsyn Institute of Nuclear, Physics, Moscow, Russia
\item \Adef{orgIII}Department of Applied Physics, Aligarh Muslim University, Aligarh, India
\item \Adef{orgIV}Institute of Theoretical Physics, University of Wroclaw, Poland
\end{Authlist}

\section*{Collaboration Institutes}
\renewcommand\theenumi{\arabic{enumi}~}
\begin{Authlist}
\item \Idef{org1}A.I. Alikhanyan National Science Laboratory (Yerevan Physics Institute) Foundation, Yerevan, Armenia
\item \Idef{org2}Benem\'{e}rita Universidad Aut\'{o}noma de Puebla, Puebla, Mexico
\item \Idef{org3}Bogolyubov Institute for Theoretical Physics, National Academy of Sciences of Ukraine, Kiev, Ukraine
\item \Idef{org4}Bose Institute, Department of Physics  and Centre for Astroparticle Physics and Space Science (CAPSS), Kolkata, India
\item \Idef{org5}Budker Institute for Nuclear Physics, Novosibirsk, Russia
\item \Idef{org6}California Polytechnic State University, San Luis Obispo, California, United States
\item \Idef{org7}Central China Normal University, Wuhan, China
\item \Idef{org8}Centre de Calcul de l'IN2P3, Villeurbanne, Lyon, France
\item \Idef{org9}Centro de Aplicaciones Tecnol\'{o}gicas y Desarrollo Nuclear (CEADEN), Havana, Cuba
\item \Idef{org10}Centro de Investigaci\'{o}n y de Estudios Avanzados (CINVESTAV), Mexico City and M\'{e}rida, Mexico
\item \Idef{org11}Centro Fermi - Museo Storico della Fisica e Centro Studi e Ricerche ``Enrico Fermi', Rome, Italy
\item \Idef{org12}Chicago State University, Chicago, Illinois, United States
\item \Idef{org13}China Institute of Atomic Energy, Beijing, China
\item \Idef{org14}Chonbuk National University, Jeonju, Republic of Korea
\item \Idef{org15}Comenius University Bratislava, Faculty of Mathematics, Physics and Informatics, Bratislava, Slovakia
\item \Idef{org16}COMSATS Institute of Information Technology (CIIT), Islamabad, Pakistan
\item \Idef{org17}Creighton University, Omaha, Nebraska, United States
\item \Idef{org18}Department of Physics, Aligarh Muslim University, Aligarh, India
\item \Idef{org19}Department of Physics, Ohio State University, Columbus, Ohio, United States
\item \Idef{org20}Department of Physics, Pusan National University, Pusan, Republic of Korea
\item \Idef{org21}Department of Physics, Sejong University, Seoul, Republic of Korea
\item \Idef{org22}Department of Physics, University of California, Berkeley, California, United States
\item \Idef{org23}Department of Physics, University of Oslo, Oslo, Norway
\item \Idef{org24}Department of Physics and Technology, University of Bergen, Bergen, Norway
\item \Idef{org25}Dipartimento di Fisica dell'Universit\`{a} 'La Sapienza' and Sezione INFN, Rome, Italy
\item \Idef{org26}Dipartimento di Fisica dell'Universit\`{a} and Sezione INFN, Cagliari, Italy
\item \Idef{org27}Dipartimento di Fisica dell'Universit\`{a} and Sezione INFN, Trieste, Italy
\item \Idef{org28}Dipartimento di Fisica dell'Universit\`{a} and Sezione INFN, Turin, Italy
\item \Idef{org29}Dipartimento di Fisica e Astronomia dell'Universit\`{a} and Sezione INFN, Bologna, Italy
\item \Idef{org30}Dipartimento di Fisica e Astronomia dell'Universit\`{a} and Sezione INFN, Catania, Italy
\item \Idef{org31}Dipartimento di Fisica e Astronomia dell'Universit\`{a} and Sezione INFN, Padova, Italy
\item \Idef{org32}Dipartimento di Fisica `E.R.~Caianiello' dell'Universit\`{a} and Gruppo Collegato INFN, Salerno, Italy
\item \Idef{org33}Dipartimento DISAT del Politecnico and Sezione INFN, Turin, Italy
\item \Idef{org34}Dipartimento di Scienze e Innovazione Tecnologica dell'Universit\`{a} del Piemonte Orientale and INFN Sezione di Torino, Alessandria, Italy
\item \Idef{org35}Dipartimento Interateneo di Fisica `M.~Merlin' and Sezione INFN, Bari, Italy
\item \Idef{org36}European Organization for Nuclear Research (CERN), Geneva, Switzerland
\item \Idef{org37}Faculty of Electrical Engineering, Mechanical Engineering and Naval Architecture, University of Split, Split, Croatia
\item \Idef{org38}Faculty of Engineering and Science, Western Norway University of Applied Sciences, Bergen, Norway
\item \Idef{org39}Faculty of Nuclear Sciences and Physical Engineering, Czech Technical University in Prague, Prague, Czech Republic
\item \Idef{org40}Faculty of Science, P.J.~\v{S}af\'{a}rik University, Ko\v{s}ice, Slovakia
\item \Idef{org41}Frankfurt Institute for Advanced Studies, Johann Wolfgang Goethe-Universit\"{a}t Frankfurt, Frankfurt, Germany
\item \Idef{org42}Gangneung-Wonju National University, Gangneung, Republic of Korea
\item \Idef{org43}Gauhati University, Department of Physics, Guwahati, India
\item \Idef{org44}Helmholtz-Institut f\"{u}r Strahlen- und Kernphysik, Rheinische Friedrich-Wilhelms-Universit\"{a}t Bonn, Bonn, Germany
\item \Idef{org45}Helsinki Institute of Physics (HIP), Helsinki, Finland
\item \Idef{org46}Hiroshima University, Hiroshima, Japan
\item \Idef{org47}Hochschule Worms, Zentrum  f\"{u}r Technologietransfer und Telekommunikation (ZTT), Worms, Germany
\item \Idef{org48}Horia Hulubei National Institute of Physics and Nuclear Engineering, Bucharest, Romania
\item \Idef{org49}Indian Institute of Technology Bombay (IIT), Mumbai, India
\item \Idef{org50}Indian Institute of Technology Indore, Indore, India
\item \Idef{org51}Indonesian Institute of Sciences, Jakarta, Indonesia
\item \Idef{org52}INFN, Laboratori Nazionali di Frascati, Frascati, Italy
\item \Idef{org53}INFN, Sezione di Bari, Bari, Italy
\item \Idef{org54}INFN, Sezione di Bologna, Bologna, Italy
\item \Idef{org55}INFN, Sezione di Cagliari, Cagliari, Italy
\item \Idef{org56}INFN, Sezione di Catania, Catania, Italy
\item \Idef{org57}INFN, Sezione di Padova, Padova, Italy
\item \Idef{org58}INFN, Sezione di Roma, Rome, Italy
\item \Idef{org59}INFN, Sezione di Torino, Turin, Italy
\item \Idef{org60}INFN, Sezione di Trieste, Trieste, Italy
\item \Idef{org61}Inha University, Incheon, Republic of Korea
\item \Idef{org62}Institut de Physique Nucl\'{e}aire d'Orsay (IPNO), Institut National de Physique Nucl\'{e}aire et de Physique des Particules (IN2P3/CNRS), Universit\'{e} de Paris-Sud, Universit\'{e} Paris-Saclay, Orsay, France
\item \Idef{org63}Institute for Nuclear Research, Academy of Sciences, Moscow, Russia
\item \Idef{org64}Institute for Subatomic Physics, Utrecht University/Nikhef, Utrecht, Netherlands
\item \Idef{org65}Institute for Theoretical and Experimental Physics, Moscow, Russia
\item \Idef{org66}Institute of Experimental Physics, Slovak Academy of Sciences, Ko\v{s}ice, Slovakia
\item \Idef{org67}Institute of Physics, Bhubaneswar, India
\item \Idef{org68}Institute of Physics of the Czech Academy of Sciences, Prague, Czech Republic
\item \Idef{org69}Institute of Space Science (ISS), Bucharest, Romania
\item \Idef{org70}Institut f\"{u}r Kernphysik, Johann Wolfgang Goethe-Universit\"{a}t Frankfurt, Frankfurt, Germany
\item \Idef{org71}Instituto de Ciencias Nucleares, Universidad Nacional Aut\'{o}noma de M\'{e}xico, Mexico City, Mexico
\item \Idef{org72}Instituto de F\'{i}sica, Universidade Federal do Rio Grande do Sul (UFRGS), Porto Alegre, Brazil
\item \Idef{org73}Instituto de F\'{\i}sica, Universidad Nacional Aut\'{o}noma de M\'{e}xico, Mexico City, Mexico
\item \Idef{org74}iThemba LABS, National Research Foundation, Somerset West, South Africa
\item \Idef{org75}Johann-Wolfgang-Goethe Universit\"{a}t Frankfurt Institut f\"{u}r Informatik, Fachbereich Informatik und Mathematik, Frankfurt, Germany
\item \Idef{org76}Joint Institute for Nuclear Research (JINR), Dubna, Russia
\item \Idef{org77}Korea Institute of Science and Technology Information, Daejeon, Republic of Korea
\item \Idef{org78}KTO Karatay University, Konya, Turkey
\item \Idef{org79}Laboratoire de Physique Subatomique et de Cosmologie, Universit\'{e} Grenoble-Alpes, CNRS-IN2P3, Grenoble, France
\item \Idef{org80}Lawrence Berkeley National Laboratory, Berkeley, California, United States
\item \Idef{org81}Lund University Department of Physics, Division of Particle Physics, Lund, Sweden
\item \Idef{org82}Nagasaki Institute of Applied Science, Nagasaki, Japan
\item \Idef{org83}Nara Women{'}s University (NWU), Nara, Japan
\item \Idef{org84}National and Kapodistrian University of Athens, School of Science, Department of Physics , Athens, Greece
\item \Idef{org85}National Centre for Nuclear Research, Warsaw, Poland
\item \Idef{org86}National Institute of Science Education and Research, HBNI, Jatni, India
\item \Idef{org87}National Nuclear Research Center, Baku, Azerbaijan
\item \Idef{org88}National Research Centre Kurchatov Institute, Moscow, Russia
\item \Idef{org89}Niels Bohr Institute, University of Copenhagen, Copenhagen, Denmark
\item \Idef{org90}Nikhef, National institute for subatomic physics, Amsterdam, Netherlands
\item \Idef{org91}NRC Kurchatov Institute IHEP, Protvino, Russia
\item \Idef{org92}NRNU Moscow Engineering Physics Institute, Moscow, Russia
\item \Idef{org93}Nuclear Physics Group, STFC Daresbury Laboratory, Daresbury, United Kingdom
\item \Idef{org94}Nuclear Physics Institute of the Czech Academy of Sciences, \v{R}e\v{z} u Prahy, Czech Republic
\item \Idef{org95}Oak Ridge National Laboratory, Oak Ridge, Tennessee, United States
\item \Idef{org96}Petersburg Nuclear Physics Institute, Gatchina, Russia
\item \Idef{org97}Physics department, Faculty of science, University of Zagreb, Zagreb, Croatia
\item \Idef{org98}Physics Department, Panjab University, Chandigarh, India
\item \Idef{org99}Physics Department, University of Jammu, Jammu, India
\item \Idef{org100}Physics Department, University of Rajasthan, Jaipur, India
\item \Idef{org101}Physikalisches Institut, Eberhard-Karls-Universit\"{a}t T\"{u}bingen, T\"{u}bingen, Germany
\item \Idef{org102}Physikalisches Institut, Ruprecht-Karls-Universit\"{a}t Heidelberg, Heidelberg, Germany
\item \Idef{org103}Physik Department, Technische Universit\"{a}t M\"{u}nchen, Munich, Germany
\item \Idef{org104}Research Division and ExtreMe Matter Institute EMMI, GSI Helmholtzzentrum f\"ur Schwerionenforschung GmbH, Darmstadt, Germany
\item \Idef{org105}Rudjer Bo\v{s}kovi\'{c} Institute, Zagreb, Croatia
\item \Idef{org106}Russian Federal Nuclear Center (VNIIEF), Sarov, Russia
\item \Idef{org107}Saha Institute of Nuclear Physics, Kolkata, India
\item \Idef{org108}School of Physics and Astronomy, University of Birmingham, Birmingham, United Kingdom
\item \Idef{org109}Secci\'{o}n F\'{\i}sica, Departamento de Ciencias, Pontificia Universidad Cat\'{o}lica del Per\'{u}, Lima, Peru
\item \Idef{org110}Shanghai Institute of Applied Physics, Shanghai, China
\item \Idef{org111}Stefan Meyer Institut f\"{u}r Subatomare Physik (SMI), Vienna, Austria
\item \Idef{org112}SUBATECH, IMT Atlantique, Universit\'{e} de Nantes, CNRS-IN2P3, Nantes, France
\item \Idef{org113}Suranaree University of Technology, Nakhon Ratchasima, Thailand
\item \Idef{org114}Technical University of Ko\v{s}ice, Ko\v{s}ice, Slovakia
\item \Idef{org115}Technische Universit\"{a}t M\"{u}nchen, Excellence Cluster 'Universe', Munich, Germany
\item \Idef{org116}The Henryk Niewodniczanski Institute of Nuclear Physics, Polish Academy of Sciences, Cracow, Poland
\item \Idef{org117}The University of Texas at Austin, Austin, Texas, United States
\item \Idef{org118}Universidad Aut\'{o}noma de Sinaloa, Culiac\'{a}n, Mexico
\item \Idef{org119}Universidade de S\~{a}o Paulo (USP), S\~{a}o Paulo, Brazil
\item \Idef{org120}Universidade Estadual de Campinas (UNICAMP), Campinas, Brazil
\item \Idef{org121}Universidade Federal do ABC, Santo Andre, Brazil
\item \Idef{org122}University College of Southeast Norway, Tonsberg, Norway
\item \Idef{org123}University of Cape Town, Cape Town, South Africa
\item \Idef{org124}University of Houston, Houston, Texas, United States
\item \Idef{org125}University of Jyv\"{a}skyl\"{a}, Jyv\"{a}skyl\"{a}, Finland
\item \Idef{org126}University of Liverpool, Department of Physics Oliver Lodge Laboratory , Liverpool, United Kingdom
\item \Idef{org127}University of Tennessee, Knoxville, Tennessee, United States
\item \Idef{org128}University of the Witwatersrand, Johannesburg, South Africa
\item \Idef{org129}University of Tokyo, Tokyo, Japan
\item \Idef{org130}University of Tsukuba, Tsukuba, Japan
\item \Idef{org131}Universit\'{e} Clermont Auvergne, CNRS/IN2P3, LPC, Clermont-Ferrand, France
\item \Idef{org132}Universit\'{e} de Lyon, Universit\'{e} Lyon 1, CNRS/IN2P3, IPN-Lyon, Villeurbanne, Lyon, France
\item \Idef{org133}Universit\'{e} de Strasbourg, CNRS, IPHC UMR 7178, F-67000 Strasbourg, France, Strasbourg, France
\item \Idef{org134} Universit\'{e} Paris-Saclay Centre d¿\'Etudes de Saclay (CEA), IRFU, Department de Physique Nucl\'{e}aire (DPhN), Saclay, France
\item \Idef{org135}Universit\`{a} degli Studi di Pavia, Pavia, Italy
\item \Idef{org136}Universit\`{a} di Brescia, Brescia, Italy
\item \Idef{org137}V.~Fock Institute for Physics, St. Petersburg State University, St. Petersburg, Russia
\item \Idef{org138}Variable Energy Cyclotron Centre, Kolkata, India
\item \Idef{org139}Warsaw University of Technology, Warsaw, Poland
\item \Idef{org140}Wayne State University, Detroit, Michigan, United States
\item \Idef{org141}Westf\"{a}lische Wilhelms-Universit\"{a}t M\"{u}nster, Institut f\"{u}r Kernphysik, M\"{u}nster, Germany
\item \Idef{org142}Wigner Research Centre for Physics, Hungarian Academy of Sciences, Budapest, Hungary
\item \Idef{org143}Yale University, New Haven, Connecticut, United States
\item \Idef{org144}Yonsei University, Seoul, Republic of Korea
\end{Authlist}
\endgroup

\end{document}